\documentclass[preprint2]{aastex}

\def\grtsim{{_ >\atop{^\sim}}}
\slugcomment{Not to appear in Nonlearned J., 45.}

\shorttitle{Structural Properties of Spiral-Spiral (S+S) pairs.}
\shortauthors{Hern\'andez-Toledo et al.}

\begin{document}

\title{The Structural Properties of Isolated Galaxies, Spiral-Spiral
Pairs, and Mergers: The Robustness of Galaxy Morphology During 
Secular Evolution}

\author{H. M. Hern\'andez-Toledo \altaffilmark{1} and V. Avila-Reese\altaffilmark{2}}
\affil{Instituto de Astronom\'{\i}a, UNAM, A.P. 70-264, 04510 M\'exico D. F., 
M\'exico}
\author{C. J. Conselice\altaffilmark{3}}
\affil{California Institute of Technology, Pasadena, CA 91125, USA}
\and
\author{I. Puerari\altaffilmark{4}}
\affil{Instituto Nacional de Astrof\'{\i}sica, Optica y Electr\'onica, Puebla, 
MD 21218, M\'exico}

\altaffiltext{1}{E-mail: hector@astroscu.unam.mx}
\altaffiltext{2}{E-mail: avila@astroscu.unam.mx}
\altaffiltext{3}{E-mail: cc@astro.caltech.edu}
\altaffiltext{4}{E-mail: puerari@inaoep.mx}

\begin{abstract}

We present a structural analysis of nearby galaxies in spiral-spiral pairs
in optical $BVRI$ bands and compare with the structures of isolated
spiral galaxies and galaxies in ongoing mergers.  We use these
comparisons to  determine how galaxy structure changes during galaxy
interactions and mergers. We analyze light concentration ($C$), asymmetry
($A$), and clumpiness ($S$) parameters, and use the projections
of $CAS$ parameter space to compare these samples. We find that the
$CAS$ parameters of paired galaxies are correlated with the
projected separations of the pair. For the widest and closest pairs,
the $CAS$ parameters tend to be similar to those of isolated and
ongoing major mergers (ULIRGs), respectively. Our results imply
that galaxy $CAS$ morphology is a robust quantity that only changes
significantly during a strong interaction or major merger. The typical
time-scale for this change in our paired sample, based on dynamical
friction arguments, is short, $\tau \approx 0.1 - 0.5$ Gyr. We find
average enhancement factors for the spiral pair asymmetries and
clumpiness values of $\sim 2.2$ and 1.5. The $S$ parameter, which is
related to star formation activity, has a moderate level of enhancement
suggesting that this activity in modern spirals depends more on internal
processes than on external conditions.

We furthermore test the statistical criterion for picking up interacting
galaxies in an automated way by using the $A-S$ projection plane. The
diversity of our spiral pair sample in the $CAS$ space suggests that
structural/SF/morphological properties of interacting galaxies change
abruptly only when the interaction becomes very strong and that the
criteria for finding galaxies involved in major mergers from Conselice
(2003) is effective.

\end{abstract}


\section{Introduction} \label{S1}

One of the major unsolved questions in modern astronomy is understanding galaxy 
formation and evolution. The hierarchical clustering scenario, motivated by the 
inflationary Cold Dark Matter (CDM) cosmology, has served as the main theoretical 
background to address this question (for a recent review see Firmani \& Avila-Reese 
2003). From the observational side, two types of information have been used: (i) 
detailed studies of local galaxies interpreted as fossil records of the formation 
process, and (ii) the rapidly increasing observational data of the universe at large 
redshifts. An important question regarding the latter item is the identification 
of properties which can be fairly compared among the galaxy populations at different 
redshifts. The correct identification and measurement of these properties is essential 
to infer empirically how the present-day galaxy populations formed.

Some optical morphological features along the Hubble-sequence have been used as the 
primary characteristic to describe local normal galaxies. Unfortunately, these features 
describe only the ``tip of the iceberg'' in galaxies. Some morphological features are 
thought to be related to transient internal dynamical and luminous (star formation) 
processes rather than to long-term fundamental physical properties of galaxies (van den 
Bergh 1998; Conselice 2003 and references therein). It has also been shown that gross 
morphological characteristics change with wavelength \citep{Blo99}, making it even more 
difficult to use them as a basis for a physical classification of galaxies. On the other 
hand, a high fraction of distant galaxies look peculiar and cannot be placed onto the 
Hubble sequence \citep{Gla95, Abr96, Con04} suggesting that there must be a physical 
cause for this change (Conselice 2003). These facts have motivated the search for 
quantitative physical indices, which would allow galaxies to be classified in different 
stages of their evolution \citep{Ber00}. In this spirit, the latter authors and Conselice 
(2003) have proposed three structural indices to distinguish galaxies at different stages 
of evolution, namely the concentration of stellar light ($C$), the asymmetry in the light 
distribution ($A$), and a measure of its clumpiness ($S$).

According to the galaxy formation picture based on the hierarchical CDM cosmogony, disk 
galaxies assembly generically inside CDM halos. The main properties of the disks are 
determined largely by the halo mass aggregation history (and its related concentration), 
and by the halo angular momentum \citep{Mo98,Fir00,Bos00,Zav03}. On the other hand, in 
the hierarchical picture, the morphological characteristics of galaxies may change 
several times during its evolution \citep{Kau93,Bau96}.  Therefore, the structure of a 
galaxy at a given time reveals its present and past formation history.  The $CAS$ 
morphology system was designed and tested to reveal the major properties of galaxies 
(Conselice 2003). The $C$ parameter is related to the scale of a galaxy (mass, size)
and its halo angular momentum, in particular the spin parameter, which is roughly constant 
in time for most of the halos. $C$ could slightly change (increase) with time depending 
on the galaxy mass accretion rate and relaxation history. Strong changes are expected 
only when dissipative major mergers happen (when a spheroid forms, for example). The $S$ 
parameter, which traces recent star formation (SF) \citep{Con03}, is expected to change 
the most with time for a given galaxy as SF occurs. The SF rate history of isolated normal disk 
galaxies is predicted to increase smoothly, on average by a factor of $1.5-4$, from $z=0$ 
to $z\sim 1-2$ and then to decrease slightly at higher redshifts \citep{Avi03}. However, 
interactions may abruptly enhance SF (Hern\'andez-Toledo et al. 2001), increasing $S$ in this 
case. Finally, the parameter $A$ is expected to change significantly during interactions and
mergers \citep{Con00a}.

The $CAS$ parameters therefore may trace out fundamental physical stages of galaxy evolution, 
as well as possible transient processes related to interactions. In the hierarchical 
scenario the rate of halo major mergers increases dramatically with redshift \citep{Got01}. 
Observations show that in the local universe only about $2-8\%$ of galaxies are in 
interacting pairs \citep{Xu91,Patton00}, but this fraction increases significantly with 
redshift as shown for a galaxy sample up to $z\sim 0.5$ \citep{Patton02}. From these data and 
assuming short merging time scales ($<0.5$ Gyr), the latter authors infer that the merging 
rate increases as $(1+z)^{2.7}$. Thus, at $z\sim 3$, the merging rate would be $\sim 40$ 
times the present one. Interestingly, at $z\sim 3$ the observed fraction of galaxies with 
asymmetric features typical of a recent merger was found to be as large as 50\% for the most 
luminous and massive sources (Conselice et al. 2003a,b). The detailed study of the local 
interacting pairs is crucial for understanding the structures of high-redshift galaxies. In 
this spirit, the aim of the present paper is to study $CAS$ parameters and correlations for a 
local sample of spirals in pairs, and compare them with the $CAS$ parameters for a reference 
sample of non-interacting galaxies and a sample of merging galaxies, namely ultra luminous infrared 
galaxies (ULIRGs).  This further allows us to determine what physical processes are required 
to change the morphologies of galaxies.  We explore how different stages of interactions 
affect the structure, morphology, and SF properties of galaxies as revealed through $CAS$ 
parameters. In a subsequent paper \citep[Paper II]{Her04}, the correlations of the $CAS$ 
parameters with other galaxy properties in pairs will be explored in order to gain insight 
on the significance of these parameters for understanding galaxy interactions and evolution.

To address these questions, we use a set of broad-band $BVRI$ observations of 66 spirals in 
interacting pairs \citep{Her01}, a set of F555W/F606W and F814W HST observations for 66 ULIRGs 
(Conselice 2003 and references therein), as well as a comparison sample of $\sim 90$ bright, 
large non-interacting local galaxies from the catalog of Frei et al. (1996).

The structure of the paper is as follows.  Section~\ref{S2} summarizes the main characteristics 
of the samples that are relevant to this study. In \S ~\ref{S2.1} an overview of the $CAS$ 
parameters and their measurements is given. In Section 3 we present the $CAS$ parameters in 
the $B$, $V$, $R$ and $I$ bands for the (S+S) pairs and compare their loci in the $CAS$ volume 
with that of the reference samples of isolated galaxies and ultraluminous infrared galaxies 
(ULIRGs). The results are discussed and interpreted in \S ~\ref{S4}. In \S ~\ref{S5} we present 
our conclusions.

\section{Observations} \label{S2}
\subsection{(S+S) Pair Sample}

\begin{figure}
\plotone{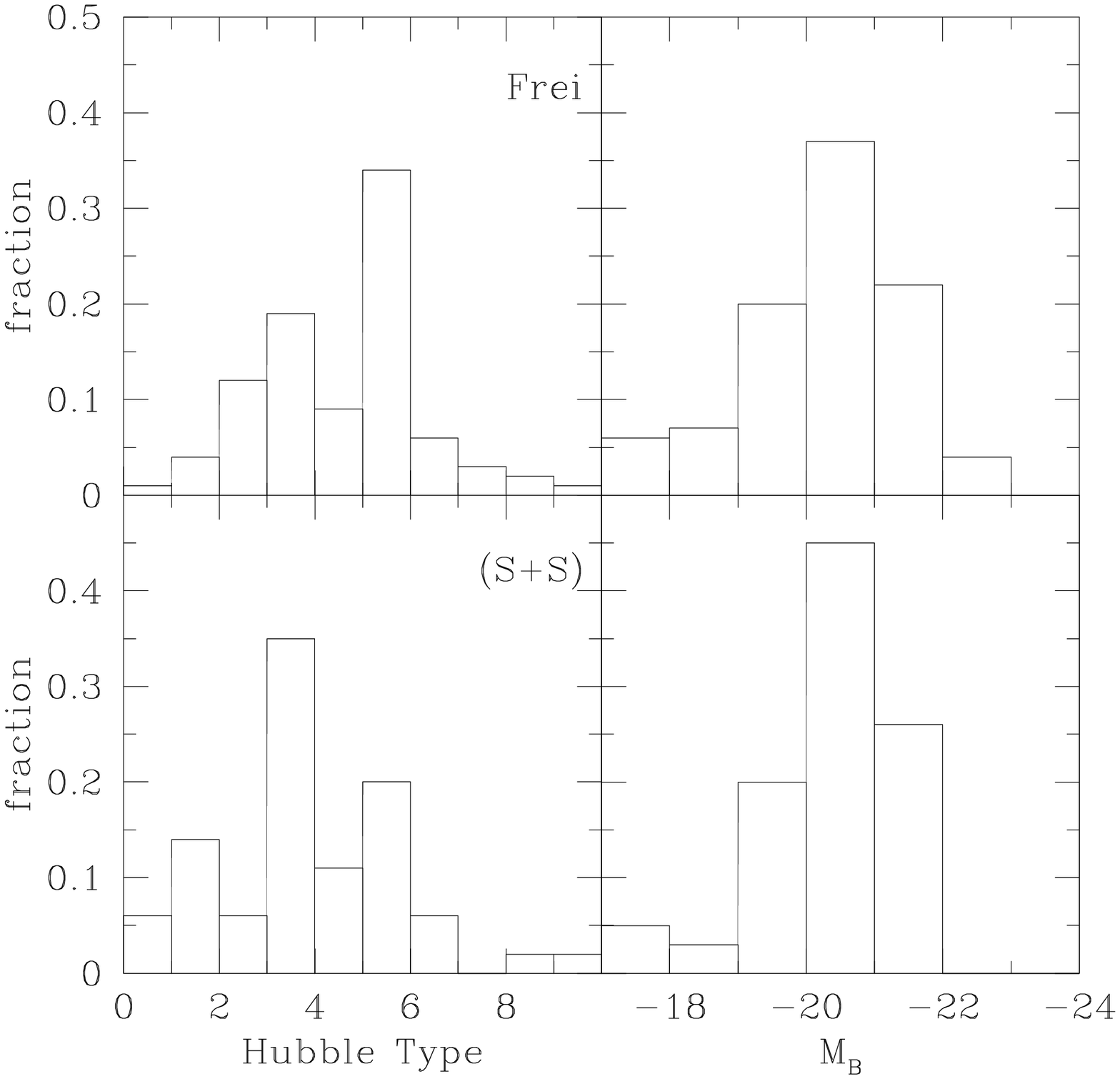}
\caption{Hubble type and blue absolute magnitude distributions in
the (S+S) pair and isolated (Frei et al.) galaxy samples. Morphological types have been 
entered as a numerical code following LEDA.\label{fig1}}
\end{figure}

The (S+S) sample used in this paper is a collection of 66 disk galaxies in pairs
observed by Hern\'andez-Toledo \& Puerari (2001) and
selected from the Catalog of Isolated Pairs of Galaxies in the Northern Hemisphere 
\citep{Kar72}. The (S+S) pairs occupy a special place among physical binaries. They 
represent a good laboratory from which to study  secular evolution due to interactions.
These pairs are also thought to be the precursors of major galaxy mergers that likely 
dominate the process through which massive galaxies form (Conselice et al. 2003a,b).  
The selection criteria applied yields a sample that is biased against the presence of 
mergers since only components with discernible diameters were selected. Instead, a wide 
range of separations and morphological features, presumably associated with tidal 
interactions, are present. The mean projected separation of the components in the sample 
is $\sim 30$ h$^{-1}_{0.7}$kpc, with the closest and widest systems separated by 
$\sim 10$ and $\sim 100$h$^{-1}_{0.7}$kpc, respectively.

The sample is restricted to an isolated environment, where only very small 
neighbors are allowed around each component. From the isolation criteria, for 
a circle of diameter equal to the  pair projected separation around each 
component, the aparent diameter of any possible neighbor can not exceed 1/5th the 
diameter of the secondary component. Smaller galaxies and/or fainter
than m$_V$= 15.7 m (the limit magnitude of the sample), might be present in 
these pairs. Although small satellite galaxies may have some influence on the 
morphological ($CAS$) properties of the pair components, one expects that
these properties in our sample are affected mainly by the effects of the mutual interactions 
between the pair components.

Hern\'andez-Toledo \& Puerari (2001) obtained deep images in the Johnson-Cousins $BVRI$ photometric 
system with typical integration times of 40, 25, 10 and 10 minutes respectively\footnote{Detailed 
imaging of these galaxy sample can be found in the VizieR On-line Data Catalog J/A+A/379/54 
associated to the Hern\'andez-Toledo \& Puerari (2001) paper}. The images were acquired using 
two telescopes: the Guillermo Haro Observatory (GHO) 2.1m telescope at Cananea, 
Sonora and the Observatorio Astron\'omico Nacional (OAN) 1.5m telescope at San Pedro M\'artir, 
Baja California, both in M\'exico and under reasonable seeing conditions ($\sim 1.6 \arcsec$).
The data are sensitive enough to detect faint stellar tidal structures. This is important in order 
of not biassing the $CAS$ estimates for a local reference sample of interacting galaxies. 
If a typical size of $\sim 1.3$ arcmin/galaxy is considered, an average number 
of 50 resolution elements/galaxy are reasonable enough to avoid underestimations of the $CAS$ 
parameters. A more detailed description of the selection criteria, completeness and 
global optical emission properties can be found in Hern\'andez-Toledo et al. (1999).

\subsection{Frei Reference Sample}

The Frei et al. (1996; hereafter (S)Frei) sample is a collection of 113 
well-resolved non-interacting galaxies of all classical Hubble types whose quantitative 
morphological properties in the $R$ band were listed in Conselice (2003, Table 1). Two 
telescopes were used to acquire these images; the Lowell 1.1m and the Palomar 1.5m with a 
typical resolution $\sim 2 \arcsec$. A typical scale of $\sim 3$ arcmin/galaxy renders an 
average number of 90 resolution elements/galaxy. This is an important requisite for a local
reference sample of non-interacting galaxies that we use for comparative purposes. The (S)Frei 
sample consists mainly of bright, high surface brightness galaxies. Hence the large population 
of low surface brightness or dwarf galaxies that make up the bulk of all galaxies in the local 
universe is under-represented. In this sense, the Frei sample is not an accurate representation 
of the entire local galaxy population but samples a wide enough range of luminosities and Hubble 
types for comparative purposes (see Bershady et al. 2000 for more details).


Figure 1 shows the morphological type and luminosity distributions of the (S+S) pair and 
(S)Frei samples.  Morphological types are represented by a numeric code, $T$, according to the 
convention in the HyperLeda database\footnote{http://leda.univ-lyon1.fr}(HyperLeda). 
Notice that only galaxies with types latter than S0 ($T\ge 0$) were considered.
The selection criteria applied to the (S+S) pairs favors the presence of bright almost 
equally-sized members with a luminosity distribution similar to that of the (S)Frei sample. 
The (S+S) pair sample also excludes, by definition, component galaxies with morphological 
types earlier than S0. The morphological distribution shows that the (S+S) sample is somewhat 
over-represented in earlier types with respect to the (S)Frei sample, where the peak in the 
distribution is in the Sc types. This excess of early types in paired galaxies with respect to 
other samples of isolated galaxies has been reported before \citep{Her99} and tentatively 
interpreted as evidence of secular or induced bulge-building.


\subsection{ULIRGs}

ULIRGs are thought to be galaxies in extreme interaction, or mergers 
\citep{Bor00,Can01}. Several groups obtained $HST$ images of ULIRGs in the 
F814W (hereafter $I$) and F555W (hereafter $V$) bands. The quantitative morphological 
properties of a compiled sample of 66 ULIRGs in these bands are presented in Conselice 
(2003). Alternative descriptions for these galaxies can be found in Ferrah et al. (2001). 
To compare the $CAS$ values properly with the $R-$band images used for the (S)Frei sample, 
a quantitative morphological $k-$correction to estimate the $R$-band value for each 
morphological index in ULIRGs was applied, as described in Conselice (2003). Considering 
a typical FWHM of $\sim 0.12 \arcsec$ and a scale of $\sim 20 \arcsec/ULIRG$, an average number 
of 160 resolution elements/ULIRG is good for $CAS$ estimates.

\subsection{$CAS$ and SEP Parameters} \label{S2.1}

In the following, we briefly review the $CAS$ parameters and discuss the 
reliability of measuring them in interacting galaxies.

\noindent {\bf Concentration of light $(C)$:}
The parameter $C$ can be measured quantitatively by using a single index that is 
calculated according to different definitions \citep{Gra01,C03}. The concentration index 
$C$, used here, is defined as the ratio of the 80\% to 20\% curve of growth radii 
($r_{80}$, $r_{20}$), within 1.5 times the Petrosian inverted radius at r($\eta = 0.2$), 
$r_P\prime$, normalized by a logarithm:
\begin{equation}
C = 5 \times log(r_{80\%}/r_{20\%}).
\end{equation}
For a detailed description of how this parameter is computed see Bershady et al. (2000) 
and Conselice (2003). The concentration of light is related to the galaxy light (or stellar 
mass) distributions. Low (high) concentrations are expected for extended (compact) galaxies 
\citep{Ber00,Gra01}.

\noindent {\bf Asymmetry $(A)$:}
The asymmetry index is defined as the number computed when a galaxy is rotated $180^{\circ}$ 
from its center and then subtracted from its pre-rotated image. A summation of the absolute 
value intensity residuals are compared with the original galaxy intensities. $A$ is also 
measured within $r_P\prime$. For a full detailed description see Conselice et al. (2000). The 
asymmetry index is sensitive to any feature that produces asymmetric light distributions. 
This includes SF, galaxy interactions/mergers, and projection effects such as dust lanes. 

\noindent {\bf Clumpiness $(S)$:}
Galaxies undergoing SF are very patchy and contain large amounts of light at high spatial 
frequency. To quantify this, the clumpiness index S is defined as the ratio of the amount 
of light contained in high frequency structures to the total amount of light in the galaxy 
within $r_P\prime$ \citep{Con03}. The $S$ parameter, because of its morphological nature, is 
also sensitive to dust lanes and inclination.

\noindent {\bf SEP:} The apparent projected separation in paired galaxies, $x_{\rm 12}$, 
is typically expressed in units of the primary component diameter, $D_{\rm 25}$, where
 $D_{\rm 25}$ is the 25-mag/arcsec$^2$ isophote diameter in the $B-$band. Thus, a 
quantity $SEP = x_{\rm 12}/ D_{\rm 25}$ 
is defined, and the (S+S) sample is sorted into wide ($SEP > 1$) and close ($SEP < 1$)
pairs. Light contamination effects are expected in paired galaxies of similar diameters 
and $SEP \lesssim 1$ or in paired galaxies with different diameters and $SEP << 1$.  If 
the light of the companion enters within the $r_P\prime$ of a given galaxy, then the observed 
$r_{P}\prime$ could be biased to a larger value, depending on the type of deformation induced 
to a ``pre-contaminated '' light profile. This will certainly affect the measured value of 
$CAS$ parameters: $C$ probably will increase (Section~\ref{S4}; Paper II), $A$, due to the 
obvious asymmetrical light contamination, will increase, and $S$ will also tend to increase 
but probably to a lesser extent. This was indeed the case for 10 galaxies in our closest 
pairs, for which a correcting procedure described below was applied.

\subsection{Measuring $CAS$ Parameters} \label{S2.2}

The measurement of the $CAS$ parameters in the S+S pairs was carried out in several 
steps: 
(i) close field and overlapping stars were removed from each image; (ii) sky background 
was also removed from the images by fitting a polynomial function that yielded the lowest 
possible residual after subtraction; (iii) the center of each galaxy was considered as the 
barycenter of the light distribution and the starting point for measurements (Conselice
2003); (iv) the $CAS$ parameters for pairs with $SEP > 1$, as well as with $SEP < 1$ but 
where the companion galaxy is beyond 2 times the $D_{\rm 25}$ of the primary, were estimated 
directly, i.e. individual components were not considered to be influenced by light 
contamination from the companion; (v) closest pairs (those with $SEP < 1$ and where the 
companion galaxy is within 1.5 times the $D_{\rm 25}$ of the primary) are considered as 
light-contaminated by the companion. In these cases, a decontamination procedure consists 
first of building a model of the pair component $a$, called $model a$, by using the task 
BMODEL in IRAF. Then we subtract this model from the original image, creating an $image 2$. 
From $image 2$, we estimated $CAS$ for component $b$. This same procedure is applied to 
component $b$ to get $model b$ in order to estimate $CAS$ for component $a$. BMODEL creates 
a model galaxy by taking into account all the information generated at the output of an 
isophotal analysis. Each model galaxy was generated in free-parameter mode, such that the 
ellipticity, the central positions ($x_{c}$,$y_{c}$) and the position angles were 
re-estimated at each galaxy radius. The resultant image, after subtracting a model galaxy 
to the original galaxy, yielded in most of the cases, traces of underlying structures. In 
such cases, these remaining features were masked or interpolated before measuring the $CAS$ 
parameters.  In one case (KPG 347), the overlapping degree makes it more than difficult to apply 
this procedure and thus we remove this pair from further analysis. The $CAS$ parameters of the 
(S)Frei and ULIRG samples (in the $R$ band) were taken from Conselice (2003).

\begin{figure}
\epsscale{1}
\plotone{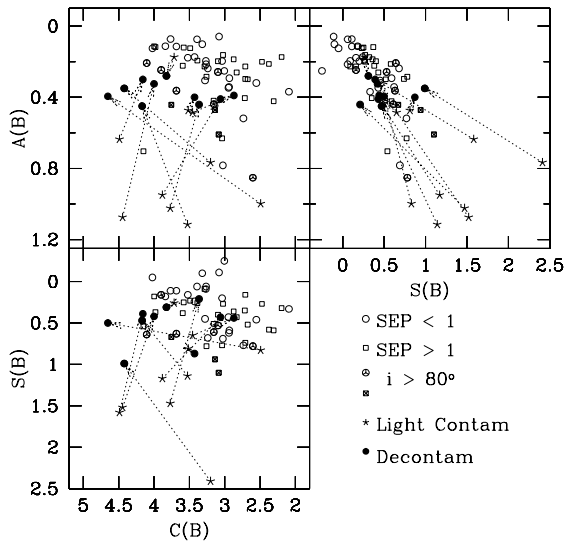}
\caption{Loci of the 66 (S+S) paired galaxies in the $B-$band $CAS$ planes.  
Different symbols are explained in the lower right panel (see text). 
The dashed arrows connect crude (asterisks) to decontaminated $CAS$ values (solid 
circles) after applying a decontamination procedure (for the closest pairs) 
described in the text.\label{fig2}}
\end{figure}

\begin{figure}
\epsscale{1}
\plotone{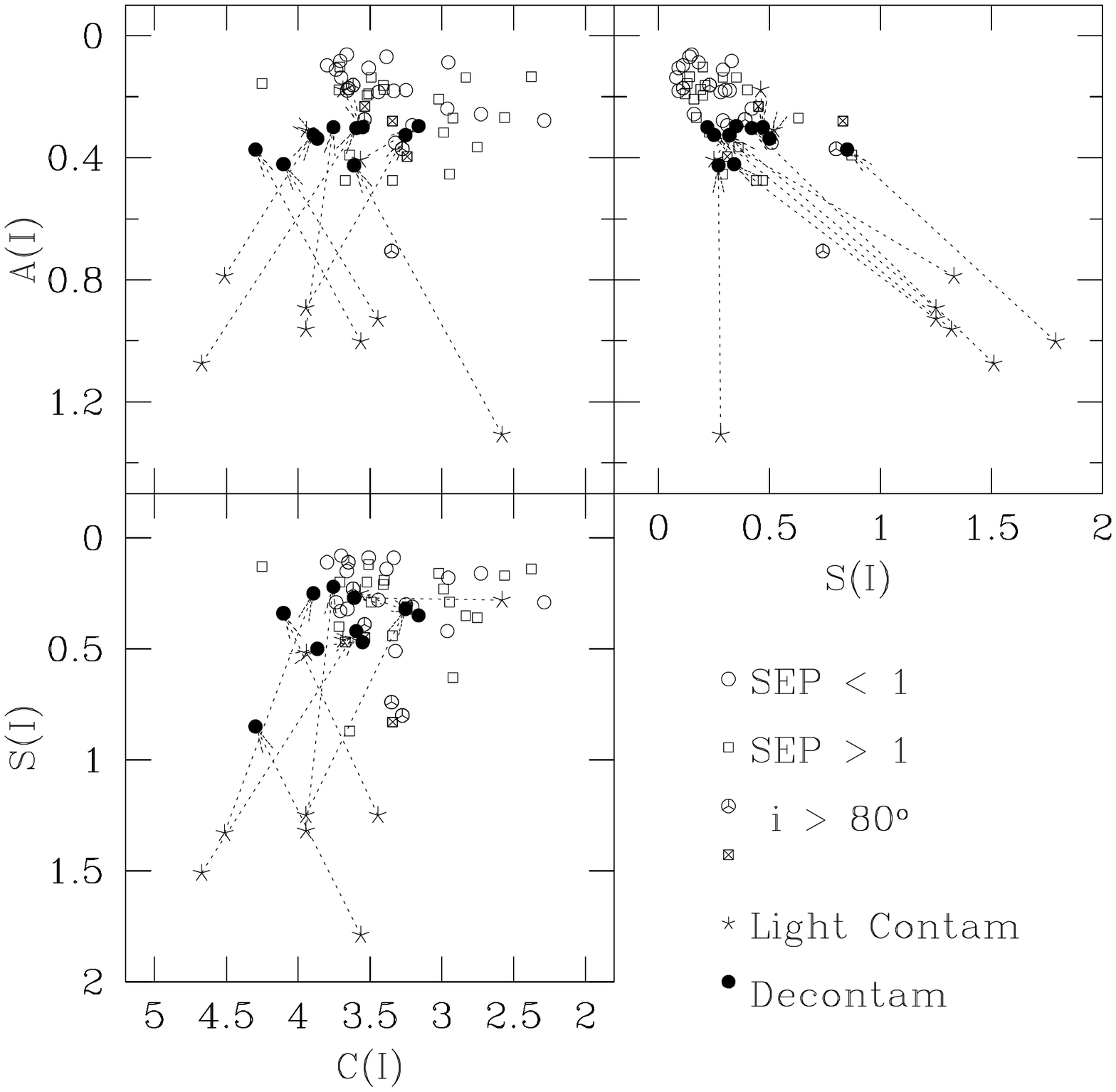}
\caption{Same as in Fig. 2 but for the $I-$band. \label{fig3}}
\end{figure}

\section{The $CAS$ Parameters of (S+S) Paired Galaxies} \label{S3}

Tables 1 and 2 list the estimated $CAS$ parameters in the $B$, $V$, $R$ and $I$ bands for 
our sample of 66 (S+S) pairs. In Figs. 2 and 3 the loci of the paired galaxies in the $A-C$, 
$A-S$, and $S-C$ planes are shown in the $B$ and $I$ bands, respectively. The data are 
presented into two groups: close paired galaxies with $SEP < 1$ (circles, mean projected 
separation $<x_{12}> \sim 23$ h$^{-1}_{0.7}$kpc) and wide pairs with $SEP > 1$ (squares, 
$<x_{12}> \sim 46$ h$^{-1}_{0.7}$kpc). The trends of these two groups in the $CAS$ space are 
slightly different. Asterisks show the crude (non-corrected) $CAS$ values for the (ten) 
closest paired galaxies. The dashed arrows connect the crude to the decontaminated $CAS$ 
values (solid circles) after applying the procedure described above. Galaxies with an 
inclination greater than $80^{\circ}$ are marked with a cross.



\subsection{Inclination Effects}\label{S3.1}

Since high inclination could introduce a systematic biased trend in the values of the $CAS$ 
parameters, it is important to evaluate its influence on the estimated parameters. Inclination 
to the line-of-sight is determined from the apparent flattening log $r_{25}$ (the axis ratio 
of the isophote 25 mag/$arcsec^{2}$ in the $B-$band taken from HyperLeda 
database and morphological type T, using the classical 
Hubble formula sin$^{2}(i) = (1 -10^{-2}$ log $r_{25}$)/($1 -10^{-2}$ log $r_{o}$), where 
log $r_{o} = 0.43 + 0.053T$ for $-5 < T < 7$ and log $r_{o} = 0.38$ for $T > 7$. In Fig. 4, 
the $B$ and $I$ band $CAS$ parameters for the (S+S) pairs are plotted vs the inclination. The 
(S)Frei sample in the $R$ band is included for comparison. A few pairs with inclinations lower 
than $20^\circ$ could be indicating a selection effect produced by interaction distortion. 
However, since the corresponding number in the (S)Frei sample is low (only five galaxies), we
expect no important bias in our comparison analysis. There is no significant correlation 
with the inclination. If any, the clumpiness parameter $(S)$ should be
slightly larger for highly inclined galaxies (see Conselice 2003). From Fig.~4 we may 
conclude that the uncertainty introduced by high inclination in the $CAS$ parameters is not 
responsible for any significant trend in $CAS$ space (Figs. 2, 3, 5 and 6). In any case, 
galaxies with inclinations larger than $80^\circ$ are marked with a skeletal triangle on 
these plots and will not be included in the statistical calculations presented below.


\subsection{Morphological Changes with Wavelength}\label{S3.2}

Table 3 presents the averages and $1 \sigma$ variations of the $CAS$ parameters in the $B, 
V, R$ and $I$ bands for the whole (S+S) sample (left to the diagonal) and close ($SEP < 1$) 
(S+S) pairs (right to the diagonal). Galaxies with $i > 80^{\circ}$, and the closest pair 
(KPG 347), were not included in the statistics (see \S 2.1). The average $CAS$ parameters in 
different optical bands show only slight differences among them. The mean values of $A$ and 
$S$ and their variances increase towards bluer bands. A conventional statistic for 
measuring the significance of a difference of means (Student's T test) between the $B$ and $I$ bands 
indicates no significant differences. Alternatively, a paired TP test that takes into account 
point-by-point effects in the two samples indicate that the $A$ parameter is significantly 
different at the 99 \% level (against the 90 \% level for the $S$ parameter). However, 
an F test for testing the hypothesis that two samples have different variances 
indicates different variances in these cases. Thus caution must be taken about the interpretation 
of the differences in means of the $A$ and $S$ parameters. The $S$ parameter, and in a lesser extent 
the $A$ parameter (e.g., Conselice 2003), are related to SF activity, which is better traced 
by the bluer bands than by the redder ones. This hints that the observed differences in $A$ and 
$S$ may be real. 

\begin{figure}
\plotone{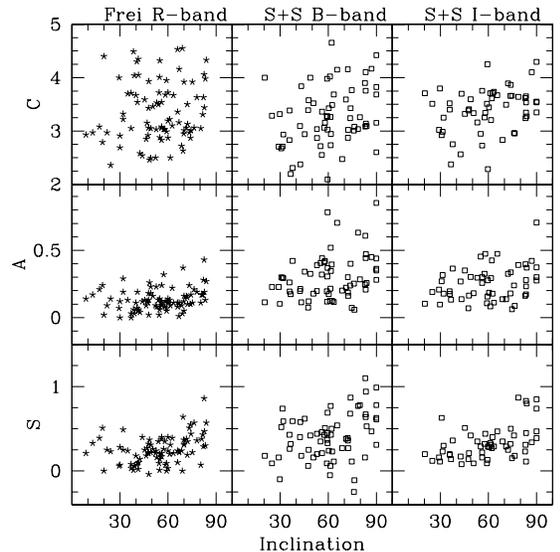}
\caption{$CAS$ parameters vs inclination for galaxies in the
isolated (Frei et al.) and paired galaxy samples. Non-significant 
trends with inclination are seen for both samples in the displayed
bands. \label{fig4}}
\end{figure}

On the other hand, the average concentration $C$ parameter decreases on average from $I$ 
toward $B$ bands. Since the samples show no significant difference in variances, a paired 
TP test supports this result at the 97 \% level. It is well known that galaxy light distributions 
are more extended (less concentrated) in bluer bands than in red bands \citep{dJ96}. However, 
it is possible that another effect for interacting galaxies, discussed in \S 4, works in the 
opposite direction, by increasing $C$ in bluer bands. The level of change 
observed in the $CAS$ parameters of interacting galaxies from $B$ to $I$ bands implies 
that the SF activity and morphological properties of disk galaxies are not strongly 
affected by the interaction, at least in periods of $\sim 1$ Gyr, which are
larger than the typical merging time scale for these interacting galaxies.

\subsection{Loci of Paired Galaxies in the Structural $CAS$ Parameter 
Space}\label{S3.3}

Figure 5 shows a comparison between (S+S) pairs and the isolated galaxy control sample in 
projections of $CAS$ space. Since the (S)Frei sample data is only available in the $R$ band, 
we carry out the comparison in this band. In order to avoid overcrowded diagrams, the 
(S)Frei data are presented in form of error bars (continuous line), corresponding to the 
average and $1-\sigma$ dispersion of their $CAS$ parameters, binned into morphological 
types: Sa-Sb, Sc-Sd and Irr, from left to right in each panel, respectively \citep{Con03}. 
The dot-dashed error bars correspond to the average and $1-\sigma$ dispersion of the $CAS$ 
parameters for a sample of five prototype starburst galaxies \citep{Con03}.

\begin{figure*}
\epsscale{2}
\plotone{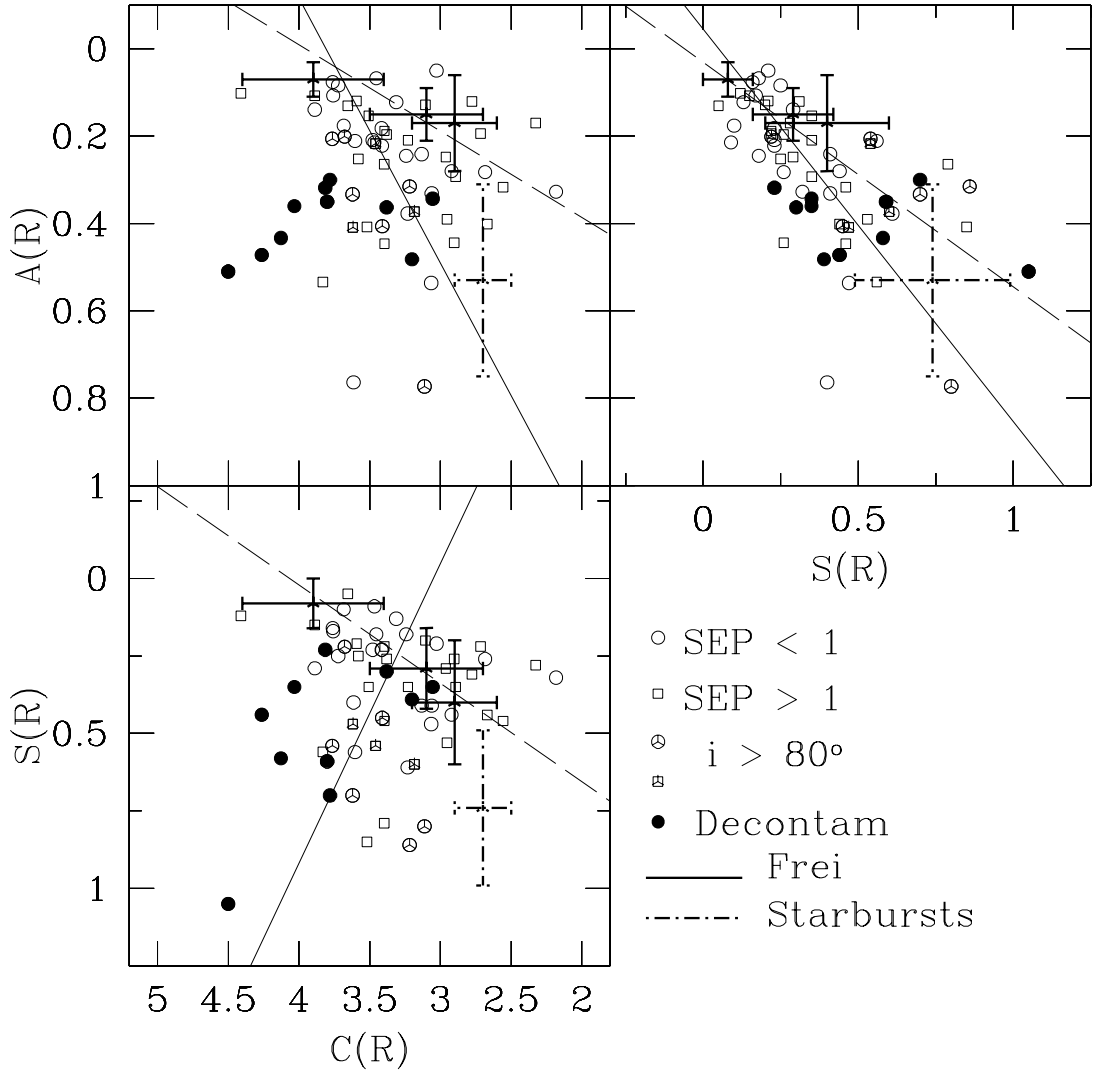}
\caption{As in Fig. 2, but in the $R-$band and including the regions
where the isolated (Frei et al.) and starbursts galaxies fall 
with their corresponding 1 $\sigma$ variations. Thick solid error bars, from
left to right in each panel, correspond to the Sa-Sb, Sc-Sd, and Irr 
isolated galaxies, respectively. Dot-dashed error bars 
correspond to the starburst sample. Solid and dashed 
lines are the bisector linear fittings to pairs with $SEP<1$ and 
isolated galaxy samples, respectively (see text) \label{fig5}}
\end{figure*}

\subsubsection{Paired vs isolated galaxies}\label{S3.3.1}

From Fig. 5 the loci of several galaxies in wide (S+S) pairs ($SEP > 1$, squares) nearly 
follow that of the isolated galaxies, while galaxies in close ($SEP < 1$, circles) pairs 
tend to deviate more from the isolated ones. This shows that galaxy structures, morphologies 
and SF activity are not strongly affected by long-term galaxy interactions or secular 
evolution. As a guideline for the eye, we have applied a linear regression fitting (bisector)
to the (S)Frei sample data (dashed line) and the (S+S) pair sample (only $SEP<1$ galaxies, 
solid line). Galaxies with $i > 80^{\circ}$ as well as the closest pair KPG347, were not 
taken into account in the fit. We used a bisector regression because we do not know a 
priori which are the physically independent variables in each diagram \citep{Isobe90}. 
Some systematic differences can be seen: for the isolated galaxies some correlation appears 
in the $A-C$, $A-S$ and $S-C$ diagrams (Pearson correlation coefficients, R, of $-0.48$, 
0.74, and $-0.60$, respectively), while for the paired galaxies a correlation is marginally
significant only in the $A-S$ diagram (R=0.68). 

Isolated galaxies exhibit a clear tendency to be more asymmetric and patchy at lower 
concentration values (and later Hubble types). Paired galaxies with $SEP > 1$ (wide pairs) 
follow a similar trend in the $A-S$ plane, although with a larger scatter. For galaxies in 
close pairs ($SEP<1$), a clearer tendency to deviate from these correlations in the $S-C$ and 
$A-C$ diagrams is observed, with several of the most concentrated pairs with large 
clumpinesses and asymmetries (see also the $B$ band results in Fig. 2). In \S \ref{S4} we 
discuss possible reasons for this behavior and what could be driving the weak structural 
evolution in interacting galaxies.

Table 4 shows the $R$ band averages and $1\pm \sigma$ variations of the $CAS$ parameters 
for (S+S) pairs, isolated galaxies, and ULIRGs, respectively. Additionally, averages and 
$1\pm \sigma$ deviations corresponding to galaxies in close and wide ($SEP < 1$ and $SEP > 1$) 
pairs  and ULIRGs in advanced stage of merging, i.e., excluding ULIRGs whose images still show 
discernible components (see below), are also presented.

 The mean values of the $A$ and $S$ parameters for close pairs ($SEP < 1$) in the $R$ band 
are $\sim 2.2$ 
and $\sim 1.5$ times larger than those for the isolated galaxies, respectively. A T test 
supports this result at the 99.9 and 99.8 \% level, respectively. A Kolmogorov-Smirnov 
test also shows different distributions in the samples for both parameters at a high significance 
level ($3.2 \times 10^{-4}$ and $1.8 \times 10^{-2}$, respectively). Thus, gravitational interactions 
tend to increase more quickly the global asymmetry than the clumpiness, consistent with asymmetries 
being more sensitive to the effect of ongoing galaxy interactions/mergers (Conselice 2003). 
The amplitude of the observed asymmetry is consistent with the results of Iono et al. 
(2004) who have recently investigated the detailed response of the $C$ and $A$ structural 
parameters for both the stellar and gaseous components in different stages of a disk-disk 
collision. They predict an increase in the stellar asymmetry of a factor $\sim 2$ over a period of 
$\sim 7 \times 10^{7}$ years that is maintained in amplitude until the end of their simulations.
These authors comment that the small dynamic range predicted for $A$ simply reflects the
intrinsic axisymmetric nature of the tidally induced features and that blindly applying the
parameter $A$ to examine the degree of tidal disturbance therefore can result in large uncertainties.
Regarding this last point, we argue that well-selected and deeply observed samples like the S+S pairs, 
with a wide range of orbital parameters and morphological features, are necessary.

On the other hand, the Student's T test shows no significant difference in the means for the 
$C$ parameter of pairs and the reference Frei sample. If we consider that 
in (S+S) paired galaxies, early Hubble types are overabundant with respect to
the reference sample (Fig. 1), and that $C$ for the reference galaxies 
increases in average as the type is earlier (Conselice 2003, Table 6), then higher 
average values of $C$ in pairs than in isolated galaxies are naively expected. 
However, galaxy interactions could affect not only the morphologycal type but also the 
concentration parameter $C$ (e.g., Iono et al. 2004), in such a way that it would
not be correct to use the $C-T$ dependence of the reference galaxies for the (S+S) 
pairs. In fact, we find that there is some trend in paired galaxies
to have higher concentrations as the parameter $SEP$ is smaller.

\subsubsection{Paired vs Ongoing Major Mergers}

Figure 6 shows a comparison of the (S+S) pairs (circles and squares) and the 66 ULIRGs 
(ongoing major mergers, dots) (\S 2) in $CAS$ structural space. Paired galaxies tend to be, 
on average, significantly less patchy, concentrated, and asymmetric than ULIRGs. The T test 
supports the differences in means for the $A$ parameter at the 96.5 \% level. However, 
an F test indicates significantly different variances in the $C$ and $S$ parameters,
complicating the interpretation of these differences. A Kolmogorov-Smirnov test however, 
supports significant differences in the distribution of the $S$ and $A$ parameters at the 
95 and 99 \% level respectively.

Notice that ULIRGs are commonly considered as individual objects, and
in consequence their $CAS$ values were measured as such (Conselice 2003).
However, from a visual examination of the HST images, we find that a 
fraction of the ULIRGs still show discernible components. Although a discussion
on whether these objects should be considered as individuals or
not is out of the scope of the present work, we identify in Fig. 6 these 
ULIRGs in process of merging with an extra cross. In Fig. 6 ULIRGs show larger 
scatter than (S+S) pairs in the three projections of the $CAS$ space. This is
supported in the $C$ and $S$ parameters by the results of the F test.  
However, an important contribution to this scatter comes namely from 
the ULIRGs with discernible components (crosses), especially in the $S$ parameter.

In the $A-S$ plane several of the highly asymmetric ULIRGs tend to be more patchy than the 
highly asymmetric paired galaxies. If only  ULIRGs with non-discernible components (ULIRGs 
in advanced merging process) are considered, then the 
differences between both samples are more dramatic in the sense that the most patchy 
``advanced'' ULIRGs are less asymmetric than the most patchy paired galaxies. The most concentrated 
``advanced'' ULIRGs tend to have average or lower than average $A$ parameters, while the 
more concentrated close (S+S) galaxies tend to have large $A$ parameters (see $A-C$ plane). 
The most patchy ``advanced'' ULIRGs tend to have average or even lower than average 
concentrations (see $S-C$ plane), opposite to the trend observed in (S+S) galaxies.


\begin{figure*}
\epsscale{2}
\plotone{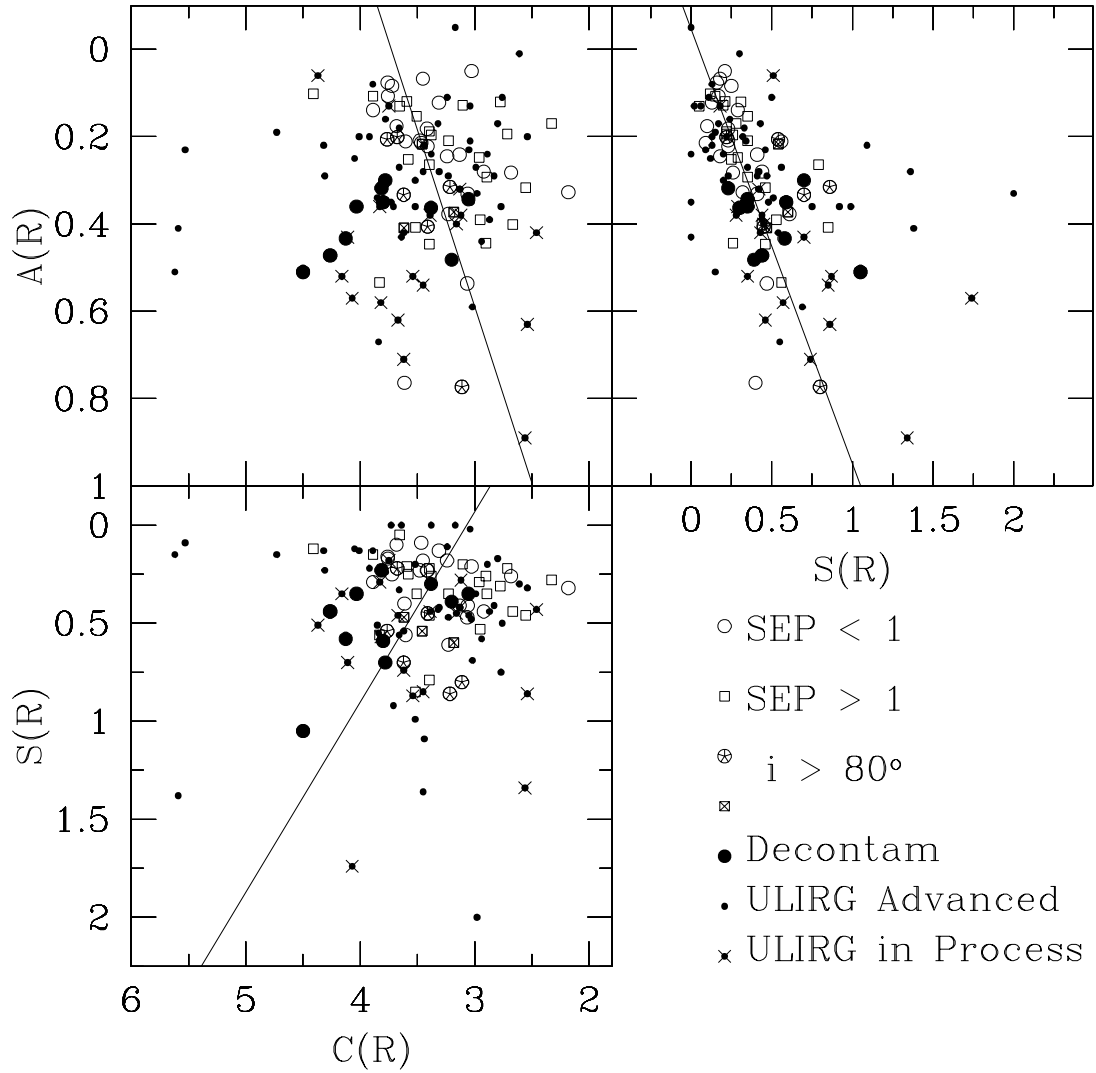}
\caption{Same as Fig. 2, but in the $R-$band and this time including  
the individual points for the ``advanced'' ULIRGs (small-solid dots) and ``in process'' 
ULIRGs (crossed small-solid dots) sample. Solid line is the bisector linear fitting to 
the pairs with $SEP<1$.
\label{fig6}}
\end{figure*}

\section{Discussion} \label{S4}

\subsection{The Robustness of Morphology in the Absence of Major 
Merging}\label{S4.1}

As we have shown, the $C$ and $A$ values for spirals in (S+S) pairs are on average larger 
than the corresponding ones to spirals in the isolated galaxy control sample. We have also shown 
that as the separation parameter $SEP$ increases, the $CAS$ values of the galaxies and 
their distribution in the $CAS$ structural space tend to be closer to those for isolated 
galaxies. On the other hand, as $SEP$ is smaller, the $CAS$ parameters and their distribution 
in the $CAS$ space become more similar to those of ULIRGs - ongoing major mergers. The large 
range of $CAS$ values for 
the (S+S) pairs reveals a large diversity of dynamical stages and morphologies in these 
systems, consistent with the expectations from their selection criteria. A representative 
projected separation and relative radial velocity (averages) can be used to estimate the 
average merging time in close ($SEP < 1$) and wide ($SEP > 1$) pairs. Alternatively, we can 
use dynamical-friction time arguments for Cold Dark Matter halos to estimate merging time
scales of the central galaxies, assuming circular orbits (Klypin et al. 1999). This method 
indicates that the average merging times for our close ($SEP < 1$, $<x_{12}> \sim 20 
$h$^{-1}_{0.7}$kpc) and open ($SEP > 1$, $<x_{12}> \sim 50$ h$^{-1}_{0.7}$kpc) (S+S) pairs 
are $\sim 0.15$ Gyr and $\sim 0.5$ Gyr, respectively. Our results show that the 
structural/SF/morphological properties of interacting galaxies change significantly only when 
the interaction becomes very strong just before the merger commences; the time scale to 
produce these changes is of the order of the merging time, roughly 0.15-0.5 Gyr.
A note of caution. These small time scales imply that the fraction of recently merged systems 
(c.f. luminous ellipticals) should be close to half the fraction of local paired galaxies. 
However, this fraction of field luminous (elliptical) galaxies seems not to be observed in the 
local optical luminosity function of field galaxies (c.f Sulentic \& Rabaca 1994; 
Hern\'andez-Toledo et al. 1999).

\begin{figure}
\epsscale{1}
\plotone{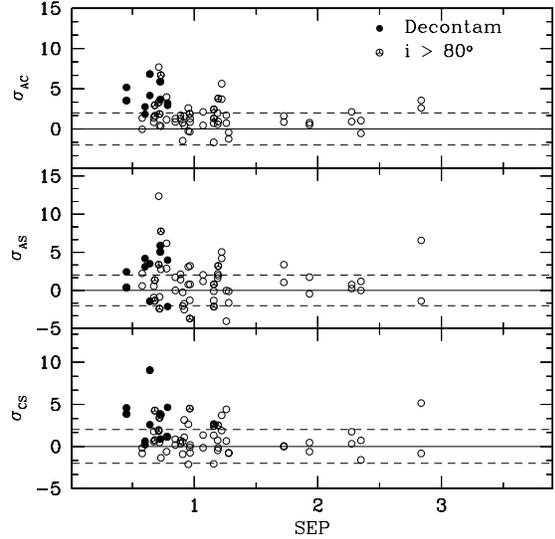}
\caption{The $\sigma$-deviations of paired galaxies from 
the isolated ones in the $R-$band $A-C$, $A-S$, and $S-C$ planes vs $SEP$,
by taking as reference a bisector linear fit to the isolated galaxies.
Dashed lines indicate the $\pm 2\sigma$ deviations. 
  \label{fig7}}
\end{figure}
 
The continuous change of the $CAS$ parameters as a function of $SEP$ is also
shown qualitatively in Fig. 7. This figure plots the $\sigma$-deviations of the (S+S) 
pairs in units of the (S)Frei sample $\sigma$-deviations in the $A-C$, $A-S$, and $S-C$ planes, 
as a function of $SEP$. Bisector linear fittings (using the corresponding formula for the variance 
$\sigma$) to the isolated (S)Frei galaxies in the $CAS$ planes were used as the reference relations. 
As the projected separation in paired galaxies decreases, the larger on average their 
$\sigma$-deviations from the isolated galaxy sample are. Dashed lines show the $\pm 2 \sigma$ 
level of the fits to the reference (S)Frei sample. In general, the largest $\sigma$-deviations 
are in the $A-S$ diagram as expected for systems that are undergoing interactions/mergers 
(Conselice 2003).

Figure~8 shows the $\sigma-$deviations for (S+S) pairs versus an interaction index introduced by 
Karachentsev (1972). The $AT$ index is designated for pairs with components in a common luminous  
halo with an amorphous, shredded, or asymmetric structure. $LI$ pairs show evidence of tidal 
bridges, tails or both in discernible components. $DI$ pairs show evidence of structural distortion 
in one or both of the separated components. Finally, a $NI$ class is introduced for wide pairs with 
no obvious morphological distortions. The order $AT-LI-DI-NI$ has been suggested 
(Hern\'andez-Toledo et al. 2001) as a sequence from strongest to weakest evidence for tidal 
distortion or, alternatively, most to least dynamically evolved (interpreting a common envelope 
as a sign of extensive dynamical evolution in pairs). The strongest $\sigma-$deviations in Fig.~8 
correspond to the $AT$ and $LI$ galaxies in the $A-S$ diagram, while pairs belonging to $DI$ and 
$NI$ classes show less than 2-$\sigma-$deviations from the reference isolated galaxies. The 
relative fraction of galaxies above 2-$\sigma-$ of the reference level are 75\%, 60\%, 40\%
and 38\% for $AT$, $LI$, $NI$ and $DI$ classes, respectively. The results in Figs. 7 and 8 are 
further evidence that the $A-S$ diagram can be used as a indicator for strong or dynamically 
evolved interactions. 


\subsection{Interacting Galaxies in the $A-S$ Plane and Induced
Star Formation}\label{S4.2}

We have discussed previously that interactions do not increase clumpiness values ($S$) as 
much as they do the asymmetries ($A$) of galaxies (see the trends for isolated and paired 
galaxies in the $A-S$ diagram of Fig. 5). According to our results, the average value of $S$ 
for close paired galaxies is $\sim 1.5$ times larger than the corresponding average for 
isolated galaxies. This enhancement factor is similar to the average one ($\sim 1.6$) in 
the optical luminosity of the late-type pair components relative to isolated galaxy 
control samples. This has been interpreted as the optical signature of the interaction-SF 
connection (e.g., Hern\'andez-Toledo et al. 1999). All these pieces of evidence show that 
while the interaction level traced by the asymmetry $A$ becomes significant, the 
interaction-induced SF rate increases only moderately in pair galaxies, suggesting that the 
SF activity in disk galaxies could be more related to internal processes rather than to 
external ones \citep{Lamb03}. The SF rate in isolated disk galaxies varies by about a factor of ten (e.g., Kennicutt 1998).

Nevertheless, when the interaction is strong and properly corresponds to the merging phase, 
the $S$ parameter increases significantly, as it is seen in the ULIRGs (Fig. 6; Conselice 2003). 
The (S+S) pairs also show a systematic increase in their $\sigma$-deviations from the reference 
isolated galaxies in the $A-S$ plane as $SEP$ decreases (see Fig. 7).  A similar trend with 
SEP was found for the 25$\mu$m-to-$B-$band (and FIR-to-$B$-band) luminosity ratio in (S+S)
pairs and (E+S) pairs (Hern\'andez-Toledo et al. 2001).

Consistent with these results, Barton et al. (2000, 2003) suggested previously that the 
strengths and the ages of triggered bursts of SF depend on the galaxy separation on the sky. 
A confrontation of their starburst models to observations indicates that the strongest bursts 
of SF occur only in the tightest orbits, giving rise to a burst strength-separation  
correlation. The burst of SF triggered by the close galaxy-galaxy pass continues and ages as 
the galaxies move apart. However, it is important to notice that the underlaying scenario of 
triggered SF in the Barton et al. (2000,2003) work is the density-driven model of SF, where the 
interaction induces a large central gas concentration. Therefore, the burst of SF occurs 
mainly in the central regions (Mihos \& Hernquist 1996). With this model it has been possible to 
produce starbursts comparable to those inferred in ULIRGs. Nevertheless large-scale, prompt and 
diverse SF activity is also seen in many interacting galaxies.

An alternative mechanism for triggered SF is the shock-induced model of SF (Jog \& Salomon 1992). 
At odds to the density-driven model, recent simulations using the shock-induced SF (Barnes 2004) 
show that (i) the SF response to the interaction is immediate, the burst being produced mainly 
at pericenter, (ii) the SF is spatially extended because the gas is more widely distributed, (iii) 
the different encounter geometry may yield a variety of SF rate histories. 
The confrontation of observations with models of spectro-photometric evolution in Barton et al. 
(2003), although applied mainly to the central regions of interacting galaxies, suggests 
indeed that the ages and strengths of triggered bursts of SF depend on the galaxy separation. 
However, in the density-driven SF model such a dependence is hard to predict since the burst is 
delayed until the gas density has had time to build up in the center, which may happen a long time 
after the pericenter pass. 

\begin{figure}
\epsscale{1}
\plotone{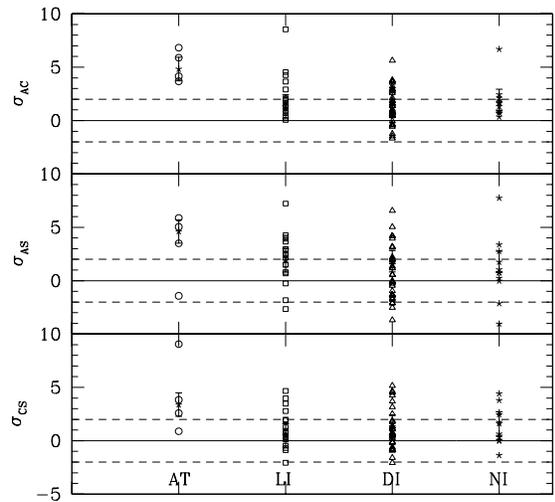}
\caption{Same as in Fig. 7, but vs the interaction class indices.
The indices AT, LI, DI and NI characterize a sequence from strongest
to weakest evidence of dynamical interaction (see text for an 
explanation of how is assigned this index to each pair galaxy). \label{fig8}}
\end{figure}

A corolary of item (ii) is that the surface brightness profiles in the bluer bands should become, 
if any, more extended than in the pre-burst galaxy, but not more concentrated. To this respect 
we have noted two trends in our paired sample: several of the close pairs tend to have a low 
average $B-$band surface brightness ($<\Sigma> = L/2/\pi r_e^2$, where $r_e$ is the effective 
radius), i.e. their $r_e$ apparently became larger. On the other hand, the $C$ parameter for 
these galaxies is typically high. Both effects can be explained together if the surface 
brightness profile increases externally, becoming shallower in the outer regions. For example, 
if an exponential surface brightness profile is gently flattened at the periphery (e.g., 
at radii $\grtsim 80\%$ of the $r_P\prime$) then the $C$ parameter used here (\S 2) will increase 
slightly while $<\Sigma>$ will decrease.

Probably, both the density-driven and the shock-induced mechanisms for SF, are working 
in galaxies, with one or the other more important in different cases, from the isolated state 
to the merging phase. Further theoretical and observational study will be important to clarify 
the mechanism of triggered SF in interacting galaxies.

In conclusion, from our study of the $CAS$ parameters for a sample of local paired galaxies we 
find that the parameter $S$, which directly correlates with SF activity (Conselice 2003), increases 
systematically with decreasing projected separation (in the plane of the sky), but this increasing 
is typically moderate.
We also find that the concentration parameter $C$ in optical bands tends 
to be high for the most asymmetric and patchy paired galaxies, opposite to the tendency of 
isolated galaxies. This could be due to either a significant increasing of the central surface 
brightness in the interacting galaxy or to a flattening at the outskirts of its surface brightness  
profile. In the former case one expects that the effective radius will decrease and $<\Sigma>$ 
will increase. This topic will be carefully analyzed elsewhere (Paper II). In the following, we 
discuss some possibilities that could produce a flattening of a surface brightness profile at 
the outskirts in an interacting galaxy.

\subsection{Concentration Parameter} \label{S4.3}

In \S 3 we have shown that some galaxies in close pairs exhibit an inverse tendency with 
respect to isolated galaxies in the $A-C$ and $S-C$ planes: their $C$ tend to be high for 
large clumpiness and asymmetry parameters (see Fig.~\ref{fig5}). These results suggest that  
close interactions may increase the concentration parameter of disks. As discussed in 
\S 4.2, the $C$ parameter is sensitive to changes in both the outer ($r>r_{80}$) 
surface brightness profile and the inner ($r<r_{20}$) one. In the former case, a 
higher value of $C$ could be due to either (a) an outer flattening of the surface 
brightness profile (see \S4.2) or (b) a dynamical expansion of the outer disk. 

In case (a), the flattening may be produced by a light contamination effect or by an induced 
physical process in a galaxy disk. Notice that we have mostly taken into account the contamination 
after applying a correction procedure described in \S 2.1.
A physical process related to interactions that could flatten the outer optical-band surface 
brightness profile is the shock-induced model of SF bursts due to interactions mentioned in 
\S 4.2. Since the gas surface density distribution of local galaxies is typically much 
more extended than the stellar surface density (and with lower values in the central 
regions), the distribution of the young stars, product of the shock-induced burst of SF, 
will also tend to be extended and with a {\it fractional} contribution to the corresponding  
(pre-burst) surface brightness more significant at the periphery than in the central regions 
(the gas surface density in local disk galaxies is typically even higher than the stellar one 
at the periphery). However, it should also be considered that differential rotation may decrease 
the molecular/neutral cloud formation in the periphery. Nevertheless, recent numerical simulations 
of shock-induced SF (instead of density-driven SF) for an observed interacting galaxy tend 
to confirm that SF is induced along the whole disk (Barnes 2004).

In case (b), Iono et al (2004) have predicted, from numerical simulations of disk-disk 
collisions, a measurable change of the parameter $C$. They have found that $C$ is 
more sensitive to the behavior of $r_{80}$ than to $r_{20}$, because for the inner 
regions, the stars respond to the collision with significantly smaller amplitudes. It 
could also be that due to the interaction the disk suffers bar instabilities and, 
in this case along with bulge formation, the outer disk expands significantly 
(Valenzuela \& Klypin 2003).  Thus, the predicted increasing of $C$ is the result of
a) the ejection of the outer stars into tidal tails and b) a rapid expansion of 
the stellar disk outside the tidal radius.

\subsection{Identifying Strong Interactions with $CAS$ Parameters}\label{S4.2}

By using a sample of ongoing major mergers (the same ULIRG sample used here), Conselice 
(2003) established a criterion for recognizing merging galaxies: 
$|A-\overline{A}_{\rm isol}|>3\sigma$, where $(A-\overline{A}_{\rm isol}$) is the residual 
of the observed value with respect to a linear fitting to the sample of isolated galaxies 
in the $A-S$ plane. In the case of pairs, according to Fig.~7 approximately 55\% (40\%) of 
the paired galaxies are above the $\pm 2 \sigma$ ($\pm 3 \sigma$) level in this plane. Two 
thirds of them have $SEP < 1$ and most of the them have clear morphological signatures of 
strong interactions (i.e., tails, bridges, distortions). In Fig.~8 one sees that most of 
the interacting galaxies (AT and LI interaction classes) are above the $\sim 2 \sigma$ level 
in the $A-S$ plane. Therefore the criterion $|A-\overline{A}_{\rm isol}|>2\sigma$ for paired 
interacting galaxies in the $A-S$ plane seems to be {\it statistically} adequate to identify 
these systems in an automated way.

On the other hand, we find that the deviations of the paired galaxies in $A-S$ space are 
slightly different to the pattern seen for ULIRGs (see Fig.~6, where ULIRGs are more scattered 
than pairs, and with a tendency to increase $S$ steeply with $A$ than the paired galaxies). 
The major differences in the $A-S$ diagram are found for ``advanced'' ULIRGs; the $A$ 
parameter in this case is relatively small for the large values of the $S$ parameter, which 
trace the SF rate. Nevertheless, even in these cases, the ULIRG SF rate traced by the optical 
$S$ parameter is much smaller than the one inferred from the FIR emission.

Finally, we notice that a merging limit criterion based only on the asymmetry parameter 
($A>0.35$ for example, Conselice et al. 2003a; Conselice, Chapman \& Windhorst 2003b) should 
be complemented with the $\sigma_{AS}-$deviation criterion when possible. In Fig. 9, 
$\sigma_{AS}$  is plotted versus $A$.  There are several paired galaxies with $A<0.35$ but 
with deviations $|A-\overline{A}_{\rm isol}|>2\sigma$, that are in fact interacting. Paired 
galaxies with $A > 0.35$, almost invariably deviate from the isolated ones by more than 
$\pm 2\sigma$.

\subsection{High Redshift Galaxy Morphology}

A goal behind the introduction of the $CAS$ system is the evaluation of the structural and 
morphological evolution of galaxies. There is mounting evidence that the population of 
unobscured star-forming galaxies has dramatically changed around $z\sim 1-2$: at higher 
redshifts, the galaxy population was dominated by peculiar galaxies, while at lower 
redshifts, most of galaxies are already ``normal'' spirals and ellipticals (Conselice, 
Blackburne \& Papovich 2004, and references therein). The latter authors have measured the 
rest-frame $B-$band $CAS$ parameters for a large sample of field galaxies in the Hubble 
Deep field North and South (HDF-N \& S) out to $z\sim 3$. The high-redshift galaxies 
($z\grtsim 2$), which are mostly peculiars, show a clear deviation from the field 
low-redshift galaxies in the $A-C$ and $A-S$ diagrams. An interesting question is whether 
the $CAS$ parameters of these high-redshift peculiars look similar to those of the local 
interacting galaxies. 

The high-redshift peculiar galaxies from Conselice et al. (2004) occupy the $A-S$ plane 
(see their Fig. 19) with values of $A$ and especially of $S$, much smaller on average than 
those of our local sample of pair (S+S) galaxies. One expects some systematic under-estimating 
of $S$ and $A$ with redshift for disks due to resolution effects (see Fig. 16 in Conselice et 
al. 2003b, and Table 1 in Conselice et al. 2004). Taking into account these effects, the loci 
of the HDF-NS peculiar galaxies with redshifts higher than $\sim 2$ in the $A-S$ plane 
become closer to the location of our local pairs. However, it is difficult to identify both 
families of objects in the $A-C$ plane as likely similar. The high-redshift peculiars have 
concentrations smaller on average than the local pair galaxies. The resolution effect for the 
high-redshift disks does not affect significantly the $C$ parameter. 
Thus, according to the $CAS$ system, a preliminary statement is that most of the HDF high-redshift 
peculiars are similar to local starburst galaxies rather than to our whole sample of local pair 
(S+S) galaxies.  Of course, there is an overlapping in the $CAS$ space of both group of galaxies 
(Fig. 5), reflecting the fact that some starburst galaxies originate in interacting systems.
It should also be taken into account that galaxies at high-redshifts are much more gas rich
than the present ones. A more careful comparison and interpretation of local and high-redshift 
galaxies based on the $CAS$ system will be presented elsewhere.

\begin{figure}
\epsscale{1}
\plotone{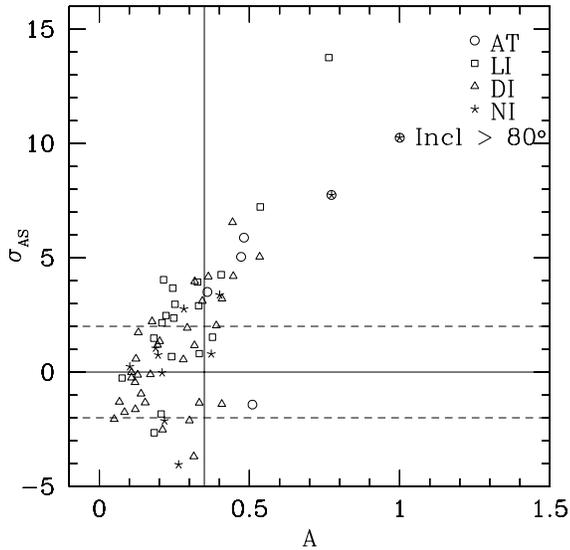}
\caption{The $\sigma-$deviations of paired galaxies from the 
isolated ones in the $A-S$ plane vs $A$. Two interaction criteria are
illustrated: $A>0.35$ (right to the solid line) and 
$|A-\overline{A}_{\rm isol}|>2\sigma$ (out from the 
dashed line region). The agreement between both criteria is 
good except for some galaxies for which $A<0.35$
but $|A-\overline{A}_{\rm isol}|>2\sigma$. These
galaxies belong in fact to the LI and DI interaction classes.  \label{fig9}}
\end{figure}

\section{Conclusions} \label{S5}

We have measured $CAS$ parameters in $BVRI$ bands for a sample of 66 galaxies in (S+S) 
pairs (Hern\'andez-Toledo \& Puerari 2001) and carried out a comparison analysis in 
two steps: 1) an inter-band internal comparison of the $CAS$ parameters in (S+S) pairs and
2) an $R$ band comparison in the $CAS$ structural space among (S+S) pairs and two samples with similarly compiled parameters, 
the (S)Frei control sample of isolated galaxies and 66 ultraluminous infrared galaxies 
\citep{Con03}. This has allowed us to gain insight on how 
morphology and SF changes as a function of interaction/mergers and recognize time-scales 
and physical effects that drive the evolution of these galaxy features. Our main conclusions 
are as follow:

\noindent (i) The differences in $CAS$ parameters for paired galaxies from the $B$ to $I$ 
bands is in general and on average small, suggesting that the optical $CAS$ parameters are 
stable galaxy properties until the beginning of a strong interaction. The average concentration 
$C$ parameter decreases on average from $I$ to $B$ bands, consistent with the finding that galaxy 
light distributions are more extended  (less concentrated) in bluer bands than in red bands 
\citep{dJ96}.

\noindent (ii) The $CAS$ parameters of paired galaxies change depending on the
separation parameter $SEP$, in the sense that the $CAS$ parameters for the widest and closest 
paired galaxies tend to be closer to those of isolated and ongoing major mergers (ULIR) galaxies, 
respectively.  
The mean values of the $A$ and $S$ parameters of the closest ($SEP <1$) paired galaxies 
are 2.2 and 1.5 times larger than those of the isolated galaxies, respectively, i.e., 
even in strongly interacting galaxies, the SF activity (related to $S$) does not 
increase too much, suggesting that this activity is more related to internal processes rather 
than external ones. 

The deviations in the $CAS$ planes of pair galaxies with respect to the isolated ones increase 
significantly on average (more than $2\sigma$) only for the closest galaxies and those galaxies 
with clear interaction signatures (Figs. 7 and 8). Nevertheless, there are also paired 
galaxies with $SEP < 1$ but within the $2\sigma$ deviations, and pairs with $SEP > 1$ but with 
deviations larger than $2\sigma$. This diversity suggests that galaxies in our sample of (S+S) 
pairs are in different dynamical and morphological stages, from non-interacting to strong 
interacting, where the projected separation $SEP$ is only an approximate (statistical) criterion 
of interaction. 

\noindent (iii) Trends in $CAS$ space for the interacting galaxies differ from trends seen 
between $CAS$ values of isolated galaxies. Within isolated systems lower concentration values 
($C$) correlate with larger $A$ and $S$ parameters. For galaxies in close pairs, these trends 
disappear and even revert for the most interacting galaxies: for several of them the larger 
$A$ and $S$, the higher is $C$.  The concentration index for strongly interacting galaxies 
may become higher owing to a ``flattening'' in the outer surface-brightness profile produced 
by interaction-induced {\it extended} SF. In the $A-S$ plane, paired galaxies, in particular 
the closer ones, increase their $A$ with increasing $S$ more steeply than isolated galaxies.

\noindent (iv) On average paired galaxies are slightly less concentrated, asymmetric, and patchy 
than ULIRGs, the significant difference being in the clumpiness. The latter display also 
a (statistically significant) larger scatter in the $CAS$ planes than the former. The major 
difference in trends between paired and ULIRG galaxies is in the $A-S$ plane: the most patchy 
ULIRGs have smaller asymmetries and slightly lower concentrations than the most patchy paired 
galaxies. The structural/SF/morphological properties of interacting galaxies change 
significantly only when the interaction becomes very strong.

\noindent (v) A criterion for picking up interacting galaxies that is based only on $A>0.35$ 
is marginally acceptable for paired galaxies. In fact there are some interacting galaxies 
with $A<0.35$.  A more robust {\it statistical} criterion uses both the $A$ and $S$ 
parameters: $|A-\overline{A}_{\rm isol}|>2\sigma$, where $(A-\overline{A}_{\rm isol})$ 
is the deviation of the interacting-galaxy $A$ parameter from the predicted one from the 
linear regression to the isolated sample in the $A-S$ diagram, and $\sigma$ is the scatter 
of this fitting.  For ongoing major mergers (ULIRGs) the $A > 0.35$ criteria is suitable.

The summary of this work is that the quantitative structural and SF parameters of galaxies 
seem to change significantly only during the strong interaction/merger phases and within 
$\sim$ 0.2-0.5 Gyr. The use of the $CAS$ parameters to establish criteria for finding galaxies 
in strong interaction at high-redshift galaxies appears to hold up as it depends strongly on 
interaction stage (e.g., Conselice et al. 2003a,b) and changes quickly with time. Within the 
local sample of (S+S) pairs analyzed here, we find the entire range of structural types are 
represented from isolated-like structural indices to indices closer to ongoing major mergers. 
This suggests that morphology is a fairly robust quantify that tends to change rapidly within 
the major merger process on the order of hundreds of Myrs.


\acknowledgements
We thank the referee for his(her) comments and suggestions that greately 
improved the manuscript. 
Partial support for this work was provided by CONACyT grant 
33776-E to V.A. and CONACyT grant 42810/A-1 to H.H-T.  
C.J.C. acknowledges support from an NSF Astronomy and Astrophysics Fellowship.



\begin{deluxetable}{rrrrrrrr} 
\tablecolumns{8} 
\tablewidth{0pc} 
\tablecaption{$CAS$ Parameters in $B$ and $V$ for (S+S) Galaxies.} 
\tablehead{ 
\colhead{}    &  \multicolumn{3}{c}{$B$-band} &   \colhead{}   & 
\multicolumn{3}{c}{$V$-band} \\ 
\cline{2-4} \cline{6-8} \\ 
\colhead{KPG} & \colhead{$M_{B}$}   & \colhead{$C$}    & \colhead{$A$} & 
\colhead{$S$}    & \colhead{$C$}   & \colhead{$A$}    & \colhead{$S$}}
\startdata

KPG64a & -21.44&  3.02$\pm$0.12& 0.78$\pm$0.01& 0.69$\pm$0.03& 3.41$\pm$0.14& 0.80$\pm$0.01& 0.86$\pm$0.03\\
KPG64b & -20.78&  3.15$\pm$0.22& 0.43$\pm$0.02& 0.61$\pm$0.03& 3.46$\pm$0.21& 0.47$\pm$0.02& 0.76$\pm$0.04\\
KPG68a & -21.40&  2.87$\pm$0.13& 0.39$\pm$0.02& 0.44$\pm$0.02& 2.97$\pm$0.10& 0.29$\pm$0.01& 0.41$\pm$0.04\\
KPG68b & -20.43&  4.65$\pm$0.09& 0.40$\pm$0.05& 0.50$\pm$0.03& 4.35$\pm$0.04& 0.29$\pm$0.02& 0.50$\pm$0.06\\
KPG75a & -20.31&  4.02$\pm$0.44& 0.12$\pm$0.00& -0.05$\pm$0.00& 3.80$\pm$0.47& 0.09$\pm$0.00& 0.25$\pm$0.00\\
KPG75b & -20.11&  3.00$\pm$0.23& 0.25$\pm$0.03& -0.25$\pm$0.01& 3.02$\pm$0.26& 0.30$\pm$0.00& 0.46$\pm$0.01\\
KPG88a & -21.04&  2.71$\pm$0.12& 0.29$\pm$0.05& 0.74$\pm$0.04& 2.82$\pm$0.12& 0.37$\pm$0.01& 0.55$\pm$0.01\\
KPG88b & -20.02&  2.37$\pm$0.21& 0.21$\pm$0.04& 0.58$\pm$0.03& 2.44$\pm$0.22& 0.29$\pm$0.01& 0.39$\pm$0.01\\
KPG98a & -20.74&  2.47$\pm$0.21& 0.20$\pm$0.04& 0.27$\pm$0.02& 2.63$\pm$0.21& 0.23$\pm$0.01& 0.39$\pm$0.01\\
KPG98b & -20.34&  3.99$\pm$0.44& 0.11$\pm$0.01& 0.18$\pm$0.01& 4.02$\pm$0.42& 0.15$\pm$0.00& 0.13$\pm$0.00\\
KPG102a& -20.66&  3.26$\pm$0.48& 0.21$\pm$0.01& 0.06$\pm$0.00& 3.54$\pm$0.54& 0.29$\pm$0.00& 0.18$\pm$0.00\\
KPG102b& -21.28&  3.03$\pm$0.26& 0.19$\pm$0.02& 0.20$\pm$0.01& 3.29$\pm$0.32& 0.21$\pm$0.00& 0.30$\pm$0.00\\
KPG103a& -20.66&  3.48$\pm$0.44& 0.12$\pm$0.03& 0.23$\pm$0.01& 3.49$\pm$0.48& 0.16$\pm$0.01& 0.17$\pm$0.00\\
KPG103b& -21.13&  3.57$\pm$0.61& 0.12$\pm$0.00& 0.25$\pm$0.00& 3.58$\pm$0.62& 0.17$\pm$0.00& 0.20$\pm$0.00\\
KPG108a& -20.28&  3.08$\pm$0.17& 0.60$\pm$0.04& 1.10$\pm$0.04& 3.08$\pm$0.19& 0.42$\pm$0.01& 0.44$\pm$0.01\\
KPG108b& -19.77&  3.13$\pm$0.11& 0.47$\pm$0.03& 0.94$\pm$0.03& 3.31$\pm$0.15& 0.28$\pm$0.01& 0.53$\pm$0.01\\
KPG112a& -19.80&  3.83$\pm$0.45& 0.28$\pm$0.00& 0.31$\pm$0.00& 3.81$\pm$0.51& 0.15$\pm$0.00& 0.19$\pm$0.00\\
KPG112b& -19.84&  4.00$\pm$0.42& 0.33$\pm$0.00& 0.42$\pm$0.01& 4.15$\pm$0.20& 0.40$\pm$0.00& 0.50$\pm$0.01\\
KPG125a& -21.02&  3.38$\pm$0.31& 0.19$\pm$0.01& 0.26$\pm$0.00& 3.46$\pm$0.31& 0.19$\pm$0.00& 0.26$\pm$0.00\\
KPG125b& -21.54&  2.30$\pm$0.13& 0.42$\pm$0.02& 0.59$\pm$0.01& 2.55$\pm$0.14& 0.40$\pm$0.00& 0.51$\pm$0.01\\
KPG136a& -21.58&  2.70$\pm$0.30& 0.22$\pm$0.01& 0.16$\pm$0.00& 2.82$\pm$0.35& 0.25$\pm$0.01& 0.20$\pm$0.00\\
KPG136b& -21.08&  3.28$\pm$0.28& 0.22$\pm$0.01& 0.09$\pm$0.00& 3.33$\pm$0.32& 0.23$\pm$0.01& 0.23$\pm$0.00\\
KPG141a& -20.46&  3.03$\pm$0.13& 0.63$\pm$0.03& 0.77$\pm$0.03& 3.45$\pm$0.15& 0.47$\pm$0.01& 1.17$\pm$0.02\\
KPG141b& -20.15&  2.71$\pm$0.28& 0.40$\pm$0.01& 0.35$\pm$0.01& 2.82$\pm$0.29& 0.41$\pm$0.00& 0.32$\pm$0.00\\
KPG150a& -20.42&  3.97$\pm$0.24& 0.11$\pm$0.01& 0.37$\pm$0.01& 4.27$\pm$0.29& 0.16$\pm$0.00& 0.16$\pm$0.00\\
KPG150b& -21.25&  2.83$\pm$0.10& 0.26$\pm$0.01& 0.43$\pm$0.01& 3.15$\pm$0.12& 0.25$\pm$0.00& 0.27$\pm$0.01\\
KPG151a& \nodata& 4.42$\pm$0.16& 0.35$\pm$0.07& 1.0 $\pm$0.04& 4.50$\pm$0.09& 0.25$\pm$0.01& 0.95$\pm$0.02\\
KPG151b& -20.79&  4.16$\pm$0.25& 0.30$\pm$0.02& 0.40$\pm$0.01& 3.97$\pm$0.17& 0.30$\pm$0.01& 0.50$\pm$0.01\\
KPG156a& -21.13&  2.97$\pm$0.10& 0.36$\pm$0.02& 0.43$\pm$0.02& 2.98$\pm$0.12& 0.31$\pm$0.01& 0.56$\pm$0.01\\
KPG156b& -19.68&  3.49$\pm$0.29& 0.20$\pm$0.01& 0.23$\pm$0.01& 3.41$\pm$0.31& 0.18$\pm$0.00& 0.31$\pm$0.00\\
KPG159a& -19.70&  2.87$\pm$0.50& 0.18$\pm$0.01& 0.36$\pm$0.80& 3.02$\pm$0.48& 0.09$\pm$0.01& 0.25$\pm$0.31\\
KPG159b& -20.51&  3.02$\pm$0.26& 0.16$\pm$0.01& 0.27$\pm$0.57& 3.26$\pm$0.29& 0.12$\pm$0.00& 0.21$\pm$0.21\\
KPG160a& -20.12&  3.39$\pm$0.21& 0.07$\pm$0.01& 0.06$\pm$0.00& 3.40$\pm$0.21& 0.07$\pm$0.00& 0.16$\pm$0.00\\
KPG160b& -19.28&  3.09$\pm$0.22& 0.25$\pm$0.02& 0.53$\pm$0.02& 3.14$\pm$0.24& 0.32$\pm$0.01& 0.79$\pm$0.02\\
KPG168a& -20.17&  3.76$\pm$0.29& 0.07$\pm$0.00& 0.11$\pm$0.00& 3.81$\pm$0.32& 0.09$\pm$0.00& 0.14$\pm$0.00\\
KPG168b& -17.85&  4.10$\pm$0.24& 0.20$\pm$0.03& 0.64$\pm$0.03& 2.04$\pm$0.02& 0.97$\pm$0.00& 0.10$\pm$0.00\\
KPG195a& -19.61&  3.89$\pm$0.37& 0.24$\pm$0.00& 0.16$\pm$0.00& 3.72$\pm$0.36& 0.22$\pm$0.00& 0.11$\pm$0.00\\
KPG195b& -19.26&  3.68$\pm$0.20& 0.36$\pm$0.02& 0.63$\pm$0.02& 3.60$\pm$0.22& 0.31$\pm$0.01& 0.59$\pm$0.02\\
KPG211a& -20.86&  3.31$\pm$0.17& 0.10$\pm$0.01& -0.10$\pm$0.00& 3.72$\pm$0.20& 0.10$\pm$0.00& 0.03$\pm$0.00\\
KPG211b& \nodata& 3.08$\pm$0.36& 0.05$\pm$0.01& -0.11$\pm$0.01& 3.17$\pm$0.38& 0.05$\pm$0.00& 0.08$\pm$0.00\\
KPG216a& -18.34&  4.15$\pm$0.14& 0.70$\pm$0.02& 0.54$\pm$0.02& 4.56$\pm$0.12& 0.74$\pm$0.01& 0.65$\pm$0.01\\
KPG216b& -19.20&  3.15$\pm$0.14& 0.42$\pm$0.01& 0.51$\pm$0.02& 3.36$\pm$0.14& 0.47$\pm$0.00& 0.49$\pm$0.01\\
KPG249a& -20.18&  3.06$\pm$0.15& 0.41$\pm$0.00& 0.43$\pm$0.01& 3.01$\pm$0.09& 0.46$\pm$0.00& 0.53$\pm$0.01\\
KPG249b& -20.35&  4.17$\pm$0.21& 0.45$\pm$0.01& 0.47$\pm$0.01& 4.08$\pm$0.07& 0.55$\pm$0.01& 0.45$\pm$0.01\\
KPG295a& -19.74&  3.36$\pm$0.17& 0.44$\pm$0.01& 0.21$\pm$0.02& 3.58$\pm$0.08& 0.35$\pm$0.01& 0.30$\pm$0.02\\
KPG295b& -20.20&  3.43$\pm$0.15& 0.40$\pm$0.01& 0.87$\pm$0.01& 3.70$\pm$0.16& 0.42$\pm$0.01& 0.79$\pm$0.01\\
KPG302a& -20.33&  2.93$\pm$0.08& 0.29$\pm$0.00& 0.59$\pm$0.01& 3.10$\pm$0.09& 0.25$\pm$0.00& 0.44$\pm$0.01\\
KPG302b& -17.24&  3.52$\pm$0.21& 0.17$\pm$0.01& 0.16$\pm$0.01& 3.52$\pm$0.19& 0.20$\pm$0.01& 0.20$\pm$0.01\\
KPG313a& -17.99&  2.19$\pm$0.08& 0.17$\pm$0.01& 0.32$\pm$0.02& 2.26$\pm$0.08& 0.17$\pm$0.00& 0.26$\pm$0.01\\
KPG313b& -18.14&  3.59$\pm$0.17& 0.22$\pm$0.00& 0.38$\pm$0.01& 3.48$\pm$0.19& 0.19$\pm$0.00& 0.40$\pm$0.00\\
KPG332a& -19.78&  2.55$\pm$0.08& 0.32$\pm$0.01& 0.46$\pm$0.01& 2.61$\pm$0.08& 0.31$\pm$0.00& 0.36$\pm$0.00\\
KPG332b& -20.15&  2.60$\pm$0.03& 0.85$\pm$0.01& 0.78$\pm$0.02& 2.92$\pm$0.03& 0.83$\pm$0.00& 0.90$\pm$0.01\\
KPG347a& -20.87&   2.69$\pm$0.07& 0.73$\pm$0.03& 0.50$\pm$0.01& 2.74$\pm$0.05& 0.50$\pm$0.00& 0.45$\pm$0.01\\
KPG347b& -21.65&   2.85$\pm$0.07& 0.49$\pm$0.04& 0.50$\pm$0.01& 2.99$\pm$0.05& 0.35$\pm$0.00& 0.50$\pm$0.01\\
KPG389a& -21.94&   2.09$\pm$0.19& 0.36$\pm$0.00& 0.33$\pm$0.01& 2.20$\pm$0.21& 0.35$\pm$0.00& 0.31$\pm$0.00\\
KPG389b& -21.49&   2.93$\pm$0.17& 0.34$\pm$0.01& 0.62$\pm$0.01& 3.13$\pm$0.19& 0.35$\pm$0.00& 0.60$\pm$0.00\\
KPG396a& -20.59&   3.75$\pm$0.13& 0.44$\pm$0.03& 0.67$\pm$0.03& 3.67$\pm$0.15& 0.46$\pm$0.01& 0.62$\pm$0.01\\
KPG396b& -19.55&   2.67$\pm$0.13& 0.29$\pm$0.01& 0.52$\pm$0.02& 2.79$\pm$0.14& 0.30$\pm$0.01& 0.42$\pm$0.01\\
KPG404a& -20.29&   3.27$\pm$0.37& 0.30$\pm$0.00& 0.45$\pm$0.01& 3.21$\pm$0.42& 0.26$\pm$0.00& 0.11$\pm$0.00\\
KPG404b& -21.97&   2.73$\pm$0.09& 0.51$\pm$0.01& 0.77$\pm$0.03& 2.95$\pm$0.10& 0.49$\pm$0.00& 0.54$\pm$0.01\\
KPG426a& -21.15&   3.26$\pm$0.22& 0.19$\pm$0.02& 0.09$\pm$0.00& 3.21$\pm$0.27& 0.14$\pm$0.00& 0.16$\pm$0.00\\
KPG426b& -20.54&   3.84$\pm$0.51& 0.11$\pm$0.00& 0.18$\pm$0.00& 3.76$\pm$0.54& 0.13$\pm$0.00& 0.12$\pm$0.00\\
KPG440a& -19.08&   3.09$\pm$0.09& 0.19$\pm$0.02& 0.40$\pm$0.02& 3.18$\pm$0.09& 0.14$\pm$0.01& 0.30$\pm$0.01\\
KPG440b& -20.79&   3.25$\pm$0.08& 0.28$\pm$0.01& 0.78$\pm$0.02& 3.32$\pm$0.08& 0.27$\pm$0.00& 0.80$\pm$0.01\\
KPG455a& -20.98&   3.68$\pm$0.30& 0.11$\pm$0.01& 0.11$\pm$0.00& 3.80$\pm$0.33& 0.09$\pm$0.00& 0.19$\pm$0.00\\
KPG455b& -21.97&   3.24$\pm$0.12& 0.23$\pm$0.03& 0.68$\pm$0.02& 3.46$\pm$0.15& 0.23$\pm$0.01& 0.62$\pm$0.01\\
\enddata
\end{deluxetable} 


\begin{deluxetable}{rrrrrrrr} 
\tablecolumns{8} 
\tablewidth{0pc} 
\tablecaption{$CAS$ Parameters in $R$ and$ I$ for (S+S) Galaxies.} 
\tablehead{ 
\colhead{}    &  \multicolumn{3}{c}{$R$-band} &   \colhead{}   & 
\multicolumn{3}{c}{$I$-band} \\ 
\cline{2-4} \cline{6-8} \\ 
\colhead{KPG} & \colhead{$SEP$}   & \colhead{$C$}    & \colhead{$A$} & 
\colhead{$S$}    & \colhead{$C$}   & \colhead{$A$}    & \colhead{$S$}}
\startdata

KPG64a&  0.712&  3.61$\pm$0.16&  0.76$\pm$0.02&  0.40$\pm$0.02&    3.80$\pm$0.17&  0.67$\pm$0.02&   0.40$\pm$0.01\\
KPG64b& \nodata&  3.41$\pm$0.26&  0.40$\pm$0.03&  0.45$\pm$0.02&    \nodata&      \nodata&      \nodata\\
KPG68a& 0.602&  3.05$\pm$0.15&  0.34$\pm$0.00&  0.35$\pm$0.02&    3.16$\pm$0.17&  0.30$\pm$0.00&  0.35$\pm$0.01\\
KPG68b& \nodata&  3.38$\pm$0.19&  0.36$\pm$0.00&  0.30$\pm$0.03&    3.61$\pm$0.19&  0.43$\pm$0.01&  0.27$\pm$0.01\\
KPG75a& 0.672&  3.88$\pm$0.48&  0.13$\pm$0.00&  0.29$\pm$0.00&    3.70$\pm$0.45&  0.08$\pm$0.00&  0.33$\pm$0.00\\
KPG75b& \nodata&  2.92$\pm$0.26&  0.28$\pm$0.00&  0.44$\pm$0.01&    2.96$\pm$0.25&  0.23$\pm$0.00&  0.42$\pm$0.01\\
KPG88a& 1.070&  2.95$\pm$0.13&  0.39$\pm$0.00&  0.53$\pm$0.01&    2.98$\pm$0.13&  0.31$\pm$0.00&  0.23$\pm$0.00\\
KPG88b& \nodata&  2.55$\pm$0.22&  0.31$\pm$0.00&  0.46$\pm$0.00&    2.56$\pm$0.22&  0.26$\pm$0.00&  0.17$\pm$0.00\\
KPG98a& 2.348&  2.71$\pm$0.22&  0.19$\pm$0.01&  0.22$\pm$0.01&    2.83$\pm$0.22&  0.13$\pm$0.02&  0.35$\pm$0.01\\
KPG98b& \nodata&  3.88$\pm$0.39&  0.10$\pm$0.00&  0.15$\pm$0.00&    3.70$\pm$0.36&  0.10$\pm$0.00&  0.20$\pm$0.00\\
KPG102a& 0.965&  3.46$\pm$0.52&  0.21$\pm$0.00&  0.09$\pm$0.00&   \nodata&      \nodata&      \nodata\\
KPG102b& \nodata&  3.41$\pm$0.34&  0.18$\pm$0.00&  0.23$\pm$0.00&   \nodata&      \nodata&      \nodata\\
KPG103a& 1.932&  3.65$\pm$0.52&  0.13$\pm$0.01&  0.05$\pm$0.00&   3.71$\pm$0.49&  0.17$\pm$0.01&  0.40$\pm$0.01\\
KPG103b& \nodata&  3.59$\pm$0.62&  0.11$\pm$0.00&  0.21$\pm$0.00&   3.40$\pm$0.57&  0.16$\pm$0.00&  0.21$\pm$0.00\\
KPG108a& 1.159&  3.18$\pm$0.20&  0.37$\pm$0.01&  0.60$\pm$0.01&   3.24$\pm$0.22&  0.39$\pm$0.01&  0.31$\pm$0.01\\
KPG108b& \nodata&  3.45$\pm$0.17&  0.21$\pm$0.00&  0.54$\pm$0.00&   3.53$\pm$0.19&  0.23$\pm$0.00&  0.45$\pm$0.01\\
KPG112a& 0.452&  3.80$\pm$0.47&  0.35$\pm$0.00&  0.59$\pm$0.00&   3.55$\pm$0.45&  0.30$\pm$0.00&  0.47$\pm$0.00\\
KPG112b& \nodata&  4.13$\pm$0.20&  0.43$\pm$0.00&  0.58$\pm$0.01&   3.60$\pm$0.46&  0.30$\pm$0.00&  0.42$\pm$0.01\\
KPG125a& 1.724&  3.39$\pm$0.31&  0.18$\pm$0.00&  0.22$\pm$0.00&   3.40$\pm$0.29&  0.17$\pm$0.00&  0.19$\pm$0.00\\
KPG125b& \nodata&  2.66$\pm$0.15&  0.40$\pm$0.00&  0.44$\pm$0.01&   2.75$\pm$0.15&  0.36$\pm$0.00&  0.36$\pm$0.00\\
KPG136a& 1.190&  2.96$\pm$0.37&  0.24$\pm$0.01&  0.29$\pm$0.00&   3.02$\pm$0.38&  0.20$\pm$0.00&  0.16$\pm$0.00\\
KPG136b& \nodata&  3.58$\pm$0.36&  0.25$\pm$0.01&  0.25$\pm$0.00&   3.51$\pm$0.35&  0.19$\pm$0.00&  0.12$\pm$0.00\\
KPG141a& 2.836&  3.51$\pm$0.16&  0.40$\pm$0.01&  0.85$\pm$0.01&   3.64$\pm$0.17&  0.39$\pm$0.00&  0.87$\pm$0.01\\
KPG141b& \nodata&  2.90$\pm$0.30&  0.44$\pm$0.00&  0.26$\pm$0.00&   2.94$\pm$0.31&  0.45$\pm$0.00&  0.29$\pm$0.00\\
KPG150a& 2.273&  4.41$\pm$0.30&  0.10$\pm$0.00&  0.12$\pm$0.00&   4.25$\pm$0.28&  0.15$\pm$0.00&  0.13$\pm$0.00\\
KPG150b& \nodata&  3.38$\pm$0.14&  0.19$\pm$0.00&  0.26$\pm$0.00&   3.52$\pm$0.15&  0.19$\pm$0.00&  0.20$\pm$0.00\\
KPG151a& 0.641& 4.50$\pm$0.17&  0.51$\pm$0.01&  1.05$\pm$0.01&   4.30$\pm$0.23&  0.37$\pm$0.01&  0.85$\pm$0.01\\
KPG151b& \nodata&  4.03$\pm$0.29&  0.36$\pm$0.01&  0.35$\pm$0.01&   3.90$\pm$0.27&  0.33$\pm$0.02&  0.25$\pm$0.01\\
KPG156a& 0.889&  3.06$\pm$0.12&  0.33$\pm$0.01&  0.41$\pm$0.00&   3.20$\pm$0.13&  0.29$\pm$0.00&  0.31$\pm$0.00\\
KPG156b& \nodata&  3.47$\pm$0.30&  0.20$\pm$0.00&  0.23$\pm$0.00&   3.44$\pm$0.30&  0.18$\pm$0.00&  0.28$\pm$0.00\\
KPG159a& 1.278&  2.77$\pm$0.48&  0.12$\pm$0.00&  0.31$\pm$0.22&   2.92$\pm$0.52&  0.11$\pm$0.00&  0.37$\pm$0.23\\
KPG159b& \nodata&  3.10$\pm$0.31&  0.12$\pm$0.00&  0.20$\pm$0.12&  \nodata&      \nodata&      \nodata\\
KPG160a& 0.963&  3.45$\pm$0.23&  0.06$\pm$0.00&  0.18$\pm$0.00&   3.38$\pm$0.22&  0.06$\pm$0.00&  0.14$\pm$0.00\\
KPG160b& \nodata&  3.21$\pm$0.27&  0.31$\pm$0.01&  0.86$\pm$0.02&   3.27$\pm$0.28&  0.37$\pm$0.01&  0.80$\pm$0.01\\
KPG168a& 0.718&  3.76$\pm$0.33&  0.07$\pm$0.00&  0.16$\pm$0.00&   3.66$\pm$0.31&  0.06$\pm$0.00&  0.15$\pm$0.00\\
KPG168b& \nodata&  3.76$\pm$0.31&  0.20$\pm$0.01&  0.54$\pm$0.02&   3.64$\pm$0.25&  0.17$\pm$0.02&  0.11$\pm$0.01\\
KPG195a& 0.680&  3.67$\pm$0.41&  0.20$\pm$0.00&  0.22$\pm$0.00&   3.61$\pm$0.40&  0.16$\pm$0.00&  0.23$\pm$0.00\\
KPG195b& \nodata&  3.62$\pm$0.25&  0.33$\pm$0.00&  0.70$\pm$0.02&   3.53$\pm$0.25&  0.27$\pm$0.01&  0.39$\pm$0.02\\
KPG211a& 0.907&  3.75$\pm$0.22&  0.10$\pm$0.00&  0.17$\pm$0.00&   3.79$\pm$0.22&  0.09$\pm$0.00&  0.11$\pm$0.00\\
KPG211b& \nodata& 3.02$\pm$0.38&  0.05$\pm$0.00&  0.21$\pm$0.00&   2.95$\pm$0.37&  0.08$\pm$0.00&  0.18$\pm$0.00\\
KPG216a& 1.222&  3.83$\pm$0.19&  0.53$\pm$0.01&  0.56$\pm$0.01&   3.67$\pm$0.19&  0.47$\pm$0.01&  0.47$\pm$0.01\\
KPG216b& \nodata&  3.39$\pm$0.15&  0.44$\pm$0.01&  0.46$\pm$0.01&   3.34$\pm$0.16&  0.47$\pm$0.01&  0.44$\pm$0.01\\
KPG249a& 0.725&  3.20$\pm$0.17&  0.48$\pm$0.01&  0.39$\pm$0.01&   3.25$\pm$0.17&  0.33$\pm$0.01&  0.32$\pm$0.01\\
KPG249b& \nodata&  4.26$\pm$0.19&  0.47$\pm$0.01&  0.44$\pm$0.01&   4.10$\pm$0.17&  0.42$\pm$0.01&  0.34$\pm$0.01\\
KPG295a& 0.783&  3.81$\pm$0.21&  0.32$\pm$0.00&  0.23$\pm$0.01&   3.76$\pm$0.19&  0.30$\pm$0.01&  0.22$\pm$0.02\\
KPG295b& \nodata&  3.78$\pm$0.19&  0.30$\pm$0.00&  0.70$\pm$0.01&   3.87$\pm$0.20&  0.34$\pm$0.00&  0.50$\pm$0.01\\
KPG302a& 0.847&  3.13$\pm$0.09&  0.24$\pm$0.00&  0.41$\pm$0.01&   3.25$\pm$0.10&  0.17$\pm$0.00&  0.26$\pm$0.00\\
KPG302b& \nodata&  3.41$\pm$0.19&  0.22$\pm$0.00&  0.23$\pm$0.01&   3.33$\pm$0.18&  0.18$\pm$0.01&  0.09$\pm$0.01\\
KPG313a& 1.158&  2.32$\pm$0.08&  0.17$\pm$0.00&  0.28$\pm$0.01&   2.37$\pm$0.08&  0.13$\pm$0.01&  0.14$\pm$0.01\\
KPG313b& \nodata&  3.50$\pm$0.20&  0.15$\pm$0.00&  0.35$\pm$0.00&   3.49$\pm$0.21&  0.13$\pm$0.00&  0.29$\pm$0.00\\
KPG332a& 0.730&  2.68$\pm$0.08&  0.28$\pm$0.00&  0.26$\pm$0.00&   2.72$\pm$0.09&  0.25$\pm$0.00&  0.16$\pm$0.00\\
KPG332b& \nodata&  3.11$\pm$0.04&  0.77$\pm$0.00&  0.80$\pm$0.01&   3.35$\pm$0.04&  0.70$\pm$0.00&  0.74$\pm$0.01\\
KPG347a& 0.371&  2.66$\pm$0.05&  0.60$\pm$0.00&  0.53$\pm$0.01&   2.77$\pm$0.04&  0.60$\pm$0.00&  0.40$\pm$0.01\\
KPG347b& \nodata&  3.26$\pm$0.06&  0.43$\pm$0.00&  0.55$\pm$0.01&   3.30$\pm$0.06&  0.40$\pm$0.00&  0.37$\pm$0.01\\
KPG389a& 0.952&  2.18$\pm$0.22&  0.32$\pm$0.00&  0.32$\pm$0.00&   2.28$\pm$0.24&  0.27$\pm$0.00&  0.29$\pm$0.00\\
KPG389b& \nodata&  3.23$\pm$0.21&  0.37$\pm$0.00&  0.61$\pm$0.00&   3.32$\pm$0.22&  0.35$\pm$0.00&  0.51$\pm$0.00\\
KPG396a& 1.193&  3.61$\pm$0.16&  0.40$\pm$0.01&  0.47$\pm$0.01&   3.34$\pm$0.20&  0.27$\pm$0.02&  0.83$\pm$0.01\\
KPG396b& \nodata&  2.89$\pm$0.14&  0.29$\pm$0.01&  0.35$\pm$0.00&   2.92$\pm$0.14&  0.27$\pm$0.02&  0.63$\pm$0.01\\
KPG404a& 0.775&  3.24$\pm$0.47&  0.24$\pm$0.00&  0.18$\pm$0.00&   3.43$\pm$0.52&  0.21$\pm$0.00&  0.21$\pm$0.00\\
KPG404b& \nodata&  3.06$\pm$0.10&  0.53$\pm$0.00&  0.47$\pm$0.01&   \nodata&      \nodata&      \nodata\\
KPG426a& 0.578&  3.31$\pm$0.30&  0.12$\pm$0.00&  0.13$\pm$0.00&   3.50$\pm$0.31&  0.10$\pm$0.01&  0.09$\pm$0.00\\
KPG426b& \nodata&  3.68$\pm$0.58&  0.17$\pm$0.00&  0.10$\pm$0.00&   3.70$\pm$0.58&  0.13$\pm$0.00&  0.08$\pm$0.00\\
KPG440a& 1.260&  3.22$\pm$0.09&  0.20$\pm$0.01&  0.35$\pm$0.01&   3.22$\pm$0.10&  0.10$\pm$0.02&  0.27$\pm$0.01\\
KPG440b& \nodata&  3.39$\pm$0.09&  0.26$\pm$0.00&  0.79$\pm$0.01&   \nodata&      \nodata&      \nodata\\
KPG455a& 0.918&  3.72$\pm$0.34&  0.08$\pm$0.00&  0.25$\pm$0.00&   3.73$\pm$0.34&  0.11$\pm$0.00&  0.29$\pm$0.00\\
KPG455b& \nodata&  3.60$\pm$0.16&  0.21$\pm$0.01&  0.56$\pm$0.01&   3.65$\pm$0.18&  0.17$\pm$0.00&  0.32$\pm$0.00\\
\enddata
\end{deluxetable} 

\clearpage

\begin{deluxetable}{lrrrrr} 
\tablewidth{0pc} 
\tablecaption{Averages and 1 $\sigma$ Variations of $CAS$ Parameters
for (S+S) Paired Galaxies.} 
\tablehead{ 
\colhead{}    &  \multicolumn{1}{c}{$B$-Band} & 
\multicolumn{1}{c}{$V$-Band} &  \multicolumn{1}{c}{$R$-Band} &
\multicolumn{1}{c}{$I$-Band} \\  
\cline{1-5} \\ 
\colhead{} & \colhead{All/SEP $<$ 1}   & \colhead{All/SEP $<$ 1}    & \colhead{All/SEP  $<$ 1} & 
\colhead{All/SEP $<$ 1} }
\startdata 

$C$ & 3.21$\pm0.53/3.31\pm$0.52 & 3.32$\pm 0.51/3.40\pm 0.45$ & 
3.33$\pm 0.45/3.40\pm $0.42 & 3.35$\pm 0.42/3.41\pm$0.40 \\
$A$ & 0.28$\pm 0.16/0.28\pm$0.16 & 0.27$\pm 0.15/0.27\pm$0.16 & 
0.26$\pm 0.15/0.27\pm$0.16 & 0.23$\pm 0.12/0.22\pm$0.11 \\
$S$ & 0.35$\pm 0.25/0.31\pm$0.28 & 0.37$\pm 0.22/0.36\pm$0.21 & 
0.33$\pm 0.17/0.32\pm$0.16 & 0.28$\pm 0.15/0.27\pm$0.12 \\
\enddata 
\end{deluxetable} 


\begin{deluxetable}{rrrrrr} 
\tablecolumns{5} 
\tablewidth{0pc} 
\tablecaption{Averages and 1 $\sigma$ Variations of $R-$band $CAS$ Parameters
for (S+S) Paired, Isolated and ULIR Galaxies} 
\tablehead{ 
\colhead{} & \colhead{(S+S) All/SEP $<$ 1/SEP $>$ 1}   & \colhead{Frei}  & \colhead{ULIRGs/ULIRGs Advanced} }
\startdata 
$C$ & 3.33$\pm 0.45/3.40\pm$0.42/3.25$\pm 0.49$ & $3.32\pm 0.55$ & $3.55\pm 0.74/3.57\pm$0.80 \\
$A$ & 0.26$\pm 0.15/0.27\pm$0.16/0.25$\pm 0.13$ & $0.12\pm 0.08$ & $0.33\pm 0.17/0.27\pm$0.13 \\
$S$ & 0.33$\pm 0.17/0.32\pm$0.16/0.34$\pm 0.20$ & $0.23\pm 0.15$ & $0.48\pm 0.40/0.42\pm$0.40 \\
\enddata 
\end{deluxetable}

\end{document}